\documentclass[useAMS]{emulateapj}
\usepackage[]{natbib}
\usepackage{graphicx}

\usepackage[colorlinks = true,
            linkcolor = blue,
                  urlcolor  = blue,
            citecolor = blue,
            anchorcolor = blue]{hyperref}

\begin{document}
\title{Optical variability in IBL S5 0716+714 during  the 2013--2015 outbursts}

 \shorttitle{Optical variability in blazar S5 0716+714}
\shortauthors{Navpreet Kaur et al.}
\altaffilmark{1}
\altaffilmark{2}

\author{Navpreet Kaur \altaffilmark{1}, Kiran S. Baliyan, S. Chandra \altaffilmark{2}, Sameer \altaffilmark{3}, \and S. Ganesh}
\affil{Astronomy \& Astrophysics Division, Physical Research Laboratory, Ahmedabad 380009, India}
\altaffiltext{1}{Indian Institute of Technology, Gandhinagar, Ahmedabad 382355, India; E-mail: navpreet@prl.res.in}
\altaffiltext{2}{North-West University, Potchefstroom 2520, South Africa}
\altaffiltext{3}{Department of Astronomy \& Astrophysics, Davey Laboratory, Pennsylvania State University, PA 16802}

\date{Under Review in AJ}\

\begin{abstract}

 With an aim to explore  optical variability at diverse timescales in BL Lac source S5 0716+714, it was observed  for 46 nights during 2013 January 14 to 2015 June 01 when it underwent two major outbursts. The observations were made using the 1.2-m  Mount Abu InfraRed Observatory telescope mounted with a CCD camera. On 29 nights, the source was monitored for more than two hours, resulting in 6256 data points in R-band, to check for the intra-night variability. Observations in B, V and I bands with 159, 214, and 177 data points, respectively, along with daily averaged R-band data are used to address inter-night and long-term variability and the color behavior of S5 0716+71. The study suggests that the source shows significant intra-night variability with a duty cycle of more than 31$\%$ and  night-to-night variations. The average brightness magnitudes in B, V, R \& I bands were found to be 14.42(0.02), 14.02(0.01), 13.22(0.01) \& 13.02(0.03), respectively, while S5 0716+714 was historically brightest with R = 11.68 mag on 2015 January 18,  indicating that source was in relatively high state during this period.  A mild bluer when brighter behavior, typical of BL Lacs,  supports the shock-in-jet model. We notice larger amplitudes of variation when the source was relatively brighter. Based on the shortest time scale of variability and causality argument, upper bound on the size of the emission region is estimated to be $9.32 \times 10^{14}$ cm and the mass of the black hole to be  5.6 $\times$ 10$^8 M_{\odot}$. 
\end{abstract}


\section{Introduction}
Blazars are a sub-class of active galactic nuclei (AGN) with their relativistic jet oriented  towards the observer's line of sight \citep{bland1979, Urry1995} leading to the Doppler boosted emission from the jet. They show extreme variability in their brightness and polarization over the time scale of  minutes to tens of years. Owing to these properties, their study serves as a tool to probe deeper into the central engine to understand the structure and emission processes in AGN. The continuum spectral energy distribution (SED) of blazars is dominated by the non-thermal emission with two broad peaks covering entire electromagnetic spectrum (EMS), ranging from radio to high energy $\gamma$-rays. The first peak  in the SED lies in the sub-mm to X-ray region and is known to be due to synchrotron process in which the relativistic electrons gyrate in a strong magnetic field present inside the jet \citep{urrymush1982} and radiate by cooling. The second, high energy peak, is understood to be due to inverse Compton scattering of low energy photons, the origin of which is not understood well.  Under the leptonic scenario  \citep[see,][ for a review]{bottcher2007}, inverse Compton scattering of the low energy photons by the relativistic electrons, which gave rise to the synchrotron emission, is responsible for the high energy peak. The seed photons which are up-scattered, could either be synchrotron photons (Self Synchrotron Compton; SSC) or external photons from the accretion disk, the broad-line region, molecular torus, cosmic microwave background etc. (external Compton; EC), or a combination of both \citep[][ and references there-in]{maraschi1994}. The exact source of these seed photons is  still an open question.  As an alternative approach, hadronic models \citep{mannheim1989} are also used to explain the high energy component in the SED \citep{Zdziarski-Bottcher2015}.
\smallskip

Blazars consist  of flat spectrum radio quasars (FSRQ) and BL Lac objects, with FSRQs differentiated from BL Lacs by the presence of broad emission lines in their spectrum (with equivalent width, EW $> 5~ \AA$; \citep{Urry1995, Laurent-Muehleisen1999}). Depending upon the frequency of the synchrotron peak in their SED, BL Lacs are further classified into three categories \citep{abdo2010} - low, intermediate and high energy peaked BL-Lacs, abbreviated as LBL, IBL and HBL, respectively. The synchrotron peak frequency, $\nu^p_{sync}$ for LBL lies below $10^{14}$Hz; for IBL, between  $10^{14}$Hz and  $10^{15}$ Hz; while for HBL $\nu^p_{sync}$ $>$ $10^{15}$ Hz.    \citet{Fossati1998} found an anti-correlation between the synchrotron peak frequency and the synchrotron peak luminosity in the blazar. Also, the Compton dominance parameter, which is the ratio of  inverse Compton peak luminosity to the synchrotron peak luminosity, decreases from the high-luminosity (FSRQs) to low-luminosity blazars (BL-Lacs). This could be due to the presence of external seed photons, from BLR or torus, leading to higher inverse-Compton luminosity \citep{Sikora-begelman-rees1994}. It was, therefore, noticed that the luminosity, degree of polarization and the $\gamma-ray$ dominance decrease from FSRQ to LBL, IBL,  and  HBL while ratio of non-thermal to thermal component and the synchrotron peak frequency increase, indicating   a blazar  sequence \citep{maraschi1994, Fossati1998, ghis-tavec2009}. 
\smallskip

Since AGN are not resolvable by any existing telescope facility, understanding their structure and emission mechanisms pose a big challenge. Blazars, which are  variable over diverse time-scales across the whole spectrum, provide a viable tool as their variability time-scales, correlated variations among multi-frequency  light-curves, color variations and SEDs  are used as probes\citep[][ and references there-in]{marscher2008, marscher2010, jorstad2010, dai2015, ciprini2007}.  
The temporal variability in blazars has been classified into three categories, namely, long-term variability (LTV) - months to years \citep{Fan2005, fan2009}, short-term variability (STV) - a few days to months, and intra-night variability (INV) or micro-variability -  a few minutes to several hours within a night \citep{wagner1995, kaur-3c2017}. Though the mechanisms responsible for  the variability remain largely unclear, long-term variability could be due to the disk perturbation/instability or structural changes in the jet, e.g., precession, bending of jet \citep{MarscherGear1985, kawaguchi1998, nottale1986, nair2005}. The STV in optical flux, including inter-night variability, could be caused by intrinsic and extrinsic processes, e.g., injection of fresh plasma in the jet, shock moving down the turbulent jet, changes in the boosting factor due to change in the viewing angle, gravitational micro-lensing etc, and sometimes results in the spectral changes \citep{ghis1997, villata2002, Hong2017}. The INV, also known as microvariability, could be due to shock compression of the plasma in the jet, shock interacting with local inhomogeneities, blob passing through quasi-stationary core, changes in the viewing angle in a jet-in-jet scenario\citep{Narayan2012} or other processes causing small scale jet turbulence \citep{MarscherGear1985, marscher2008, chandra2011,kaur-3c2017}. However, the exact processes responsible for variability, in particular INV, are not well understood and significant amount of work is required to have a better understanding of this complex phenomenon. 
\smallskip 
% Mazin et al 2009 (Conf proceeding): S5  0716+714 was  detected several times  at  different flux levels by the EGRET detector on board the Comp- ton   Gamma-ray   Observatory   [Hartman et al.(1999)]. Also  AGILE  reported  detection  of  a  variable ?-ray flux  with  peak  flux  density  above  the  maximum  re- ported  from  EGRET  [Chen et al.(2008)].  Observations at   VHE?-ray   energies  by   HEGRA   resulted   in   anupper  limit  of  F(>1.6 TeV) = 3.13photons/cm2/s [Aharonian et al.(2004)].  

The intermediate BL Lac object S5 0716+714 is one of the most active blazars and makes a perfect candidate for variability study on the blazars at diverse time scales \citep{aliu2012}. It is available in the sky for a  longer time during the night (due to its high declination), is almost always active, fairly bright, and hence can be observed with moderate facilities.  It was discovered by \citet{kuhr1981} in NRAO 5 GHz radio survey with flux larger than 1 Jy
 \footnote{1 Jy = 10$^{-23}$ erg cm$^{-2}$ s$^{-1}$ Hz$^{-1}$} 
and due to its featureless spectra \citep{bierman1981}, was categorized as a BL-Lac source. \citet{nilson2008} derived a redshift of 0.31$\pm$0.08 by taking the host galaxy as a standard candle, but recently \citet{danforth2013} put a statistical upper limit of z $<$ 0.322 (with 99$\%$ confidence) on its redshift.  The source S5 0716+714 has been observed across the EMS, including its discovery as a TeV candidate in 2008 by MAGIC collaboration \citep{Anderhub2009}, when a strong optical and $\gamma-$ray correlated activity was noticed.

S5 0716+714 shows high duty cycle of variation (DCV) as reported by \citet[][ and references there-in]{chandra2011}.  Due to all these properties,  it has been the target of several multi-wavelength campaigns around the globe \citep{wagner1995, villata2002, raiteri2005, nesci2005, montagni2006, gupta2008, dai2015}, focusing on INV and STV. After being reported in its high phase, the object was followed  by \citet{bachev2012} who claimed  historical  maxima and minima of 12.08 (MJD 56194) and 13.32 (MJD 56195) in R-band, respectively. \citet{rani2013} found the $\gamma-$ray emission to be correlated with optical and radio, supporting SSC mechanism responsible for the high energy emission. However, an orphan flare in X-rays indicated to the limitation of such simple scenario. 

Investigating the long-term variability trend, \citet[][ and references there-in]{nesci2005} reported a decreasing average brightness of the source during 1961-1983 followed by an increasing one upto 2003, superposed with short term flares. They extracted source brightness data from photographic plates obtained  from the Asiago Observatory, POSS1 and Quick V surveys dating back to 1953 to generate long-term light curves. It underlined the importance of the astronomical data, even if taken for some other purpose.  Based on these data, they even predicted a decrease in the mean brightness of the source during the next 10 years, i.e., after 2003.  Indeed,  the source was inferred, from the 2003 to 2014 optical data, to  be in decreasing brightness phase  by \citet{Chandra2013th, Baliyan2016} and the present work,  suggesting a precessing jet with increasing viewing angle.

The blazar S5 0716+714 has undergone several optical outbursts in the past, superposed on the mean decreasing or increasing long-term trends as  reported by many workers \citep{Raiteri2003, nesci2005, gupta2008, larionov2013}. Micro variability (INV) on the time-scales of a few hours to 15 minutes is reported  \citep[][ and references there-in]{chandra2011, rani2013, man2016} with S5 0716+714 showing bluer-when brighter (BWB) behaviour in general. On the other hand, \citet{Raiteri2003} found a weak correlation with color, while  others did not find any  correlation between color and brightness\citep{stalin2009, agarwal2016, wu2005}. The blazar S5 0716+714 has also been reported to show (quasi-) periodic variations (QPV) in optical at several epochs and at many time scales ranging from sub-hours to years \citep{Raiteri2003, gupta2008}. However,  \citet{bhatta2016}  did not find  3 and 5 hr QPV as genuine. Recently, \citet{Hong2018} reported 50 min QPV when the source was relatively fainter during 2005 - 2012, ascribing it to the activity in the innermost orbit of the accretion disk.

The blazar  S5 0716+714 was reported achieving new historical brightness levels (11.68 in R-band) in optical on 2015 January 18 by \citet{atel2015, chandra2015}, reassuring that it will  never stop to surprise us. It, therefore, justifies a  continuous coverage of the source to help us understand the nature of blazars in general and S5 0716+714 in particular. Keeping this objective in mind and to understand the  variability characteristics, chromatic behaviour and relationship between variability amplitude and brightness of the source, here we present our results obtained from the observations during January, 2013 to June, 2015. Section 2 describes the observations and data analysis; section 3 presents the results and discussions while section 4 summarizes the work. 

\section{Observations and Data Reduction/Analysis}
To investigate intra-night and inter-night variability in BL-Lac source S5 0716+714, we carried out  optical observations using the 1.2m telescope of the  Mount Abu Infrared Observatory(MIRO), operated by the Physical Research Laboratory, Ahmedabad. The observatory is located at Gurushikhar mountain peak, about 1680 m above the sea-level, in Mount Abu (Rajasthan), India, with  a typical  seeing of 1.2 arcsec.  
The observations were taken  with liquid-nitrogen cooled Pixcellent CCD camera as the backend instrument, equipped with Johnson-Cousins optical BVRI filter set. The dimension of the CCD array is 1296 x 1152 pixels of size 22 $\mu$m each. The field of view (FOV) is about 6.5 x 5.5 arcmin$^{2}$ with a plate scale of 0.29 arcsec/pixel. The CCD readout time is 13 seconds with readout noise of four electrons and negligible dark current when cooled to a temperature of about -120$^{\circ}$.  

\smallskip
In order to study INV (microvariability), as a strategy, we monitored the source for a minimum of two hours in the Johnsons R-band with a high temporal resolution (less than a minute) to resolve any rapid flare, while for STV and LTV in the source brightness and color, 4/5 images were taken in B, V and  I-bands everyday during the campaign period.  The source and its comparison/control stars, as they appear in the finding chart available at the web-page of the Heidelberg University\footnote{\url{http://www.lsw.uni-heidelberg.de/projects/extragalactic/charts/0716+714.html}} \label{comp},-- having brightness close to that of the source  \citep{howell1986}  were kept in the same observed frame. Differential photometry was performed to minimize the effect of non-photometric conditions (however,  majority of the observations were made during photometric nights), like minor fluctuations due to turbulent sky and other seeing effects. The exposure time was decided by keeping the counts   well below the saturation limit and  in the linear regime of our CCD \citep{cellone2008}. Several twilight flat field images  and bias images were taken on each observation night to calibrate the science images. By following the above mentioned strategy, a total of 6256, 159, 214 and 177 images in R, B, V and I-band, respectively, were obtained and subjected to analysis.

The observed data were checked for  spurious features, if any, and reduced using IRAF\footnote{IRAF-Image Reduction and Analysis Facility is data reduction and analysis package by NOAO, Tuscon, Arizona operated by AURA, under agreement with NSF.} standard tasks- bias subtraction, flat fielding, cosmic ray treatment etc. The comparison stars 5 and 6\footnote{Stars taken from the sequence A,B,C,D by \citet{ghis1997} and corresponding sequence,  2, 3, 5, 6 by \citet{villata1998}}, present in the source field  were chosen to perform differential photometry . Other stars (stars 2 and 3) in the field were too bright to be used for differential photometry as they could introduce errors  (from differential photon statistics and random noise, like sky)  \citep{hwm1988}. An optimum aperture size, three times the FWHM, was used based on the prescription by \citet{cellone2008}, as a smaller apertures can give better Signal-to-Noise(SNR), but might lead to  spurious variations if the seeing was not good, while  a larger aperture  would have significant contribution from the host galaxy  thermal emission\citep{Cellone2000} and might suppress the genuine variations in the blazar flux. Aperture photometry on the blazar S5 0716+714 and comparison stars 5 \& 6, using the same aperture size, was performed using \textit{DAOPHOT} package in IRAF on photometric nights.

The aperture photometry technique was employed on a total of 6806 images in BVRI-bands and the source magnitudes were calibrated with the average magnitude of the comparison stars 5 and 6, which were also used to check for the stability of  sky during  observations, as described in equation 1 and equation 2. No correction for host galaxy of S5 0716+714 was applied as the host galaxy is much  fainter  with R-band $>$20 mag \citep{montagni2006}) than the central bright source. The differential light curves ($LC$s) were constructed to detect  INV, while BVI \& R band long-term $LC$s were generated from daily averaged values in each band. To quantify the INV nights, we applied several statistical tests, for example confidence parameter test (C$-$test), amplitude of variability ($A_{var}$) test, as discussed in the next section.

\section{Results and discussion}
As already mentioned earlier, the  photometric data obtained after the aperture photometry were used to plot the intra-night and inter-night light curves. Though the lightcurves themselves are not sufficient to reveal the complexities of the variability and blazar phenomena, they are good indicators of the emission mechanism and can help put constraints on various models. The nature of most of the light curves  differs from one night to the other,  indicating to the  emission from random and turbulent process in the jet. Since the physical mechanisms which trigger blazar variability, especially on intra-night time scales, are still debatable, any detailed study of $LC$s should add to our understanding.

 In order to identify and characterize  the nights showing INV, we performed  variability amplitude  and  confidence parameter (C-test)  tests. In the following we also discuss STV,  LTV  and  color behaviour of the source during the period of our observing campaign.

\subsection{Intra-night variability}
Blazars show rapid variability which can sample very compact sizes of their emission regions. To determine the number of  INV nights, we first excluded the  not-so-photometric nights when sky conditions were changing drastically, and those with less than two hours of monitoring. We were left with  29 nights that qualified this criterion during 2013 January - 2015 June. 
 The  $LC$s  for S5 0716+714, being very complex  with a number of features, made it very difficult to infer INV from just visual inspection, barring a few clear cases.  To resolve this problem, following statistical methods are used to quantify the INV. 

\noindent
\subparagraph{\bf Confidence parameter test (C$-$test):}
\smallskip
The C-test was first introduced by \citet{jang1997} and further generalized by \citet{Romero1999}. It is basically a ratio between calibrated source magnitudes and the differential magnitude of the comparison stars, given as,

\begin{equation}
\label{eq3}
C= \frac{\sigma_{S-C_{5,6}}}{\sigma_{C6-C5}}
\end{equation}
where, $C_{5,6}$ is the average of difference in instrumental and standard magnitudes of stars 5 and 6, $\sigma_{S-C_{5,6}}$ and $\sigma_{C6-C5}$ are the standard deviations of differential LCs.  We consider the source to be variable when confidence  parameter is greater than 2.57 (i.e., C $>$ 2.57) for more than 3$\sigma$ confidence (or 99 $\%$ confidence level)\citep{jang1997}. The standard deviation $\sigma$ for differential light curves is given by,

\begin{equation}
\label{eq3a}
\sigma = \sqrt{\frac {\sum{(m_i-\overline{m})^2}}{N-1}}
\end{equation}
where, $m_i = (m_2-m_1)_i$ is the differential magnitude of the two objects, $\overline{m} = \overline{m_2-m_1}$ represents differential magnitude averaged over the night and N is the total number of data points.

\subparagraph{\bf F-test:} 
\smallskip

F-test, also known as Fisher-Snedecor distribution test, measures the sample variances of two quantities i.e, variance of calibrated source magnitudes and that of the differential magnitudes of the comparison stars. To test the significance of variability during each night, it is written as,
\begin{equation}
 \label{eq.ftest}
 F = \frac {{{\sigma}^2}_{B}}{{{\sigma}^2}_{CC}}
\end{equation}
where ${{\sigma}^2}_{B}$, ${{\sigma}^2}_{CC}$ are the variances in the blazar magnitudes and differential magnitudes of the standard stars for nightly observations, respectively. An F-value of $\geq 3$ implies variability with a significance of more than 90\% while an  F-value of $\geq 5$ corresponds to 99\% significance level.

\noindent
\subparagraph{\bf Amplitude of variability ($A_{var}$):} 
\smallskip

The  intra-night amplitude of variability in the source is calculated  by using the expression given by \citet{heidt1996},
\begin{equation}
\label{eq4}
A_{var} = \sqrt{(A_{max}-A_{min})^{2}-2 \sigma^{2}}
\end{equation}
where, $A_{max}$ and $A_{min}$ are the maximum and minimum magnitudes in the intra-night calibrated  light curve of the source and $\sigma$ is the standard deviation in the measurement. For a night to be considered as variable, $A_{var}$ should be more than 5\%.

\subsubsection{\bf INV light curves and duty cycle of variation (DCV)}
 After performing statistical tests on the entire data set, we get 9 out of a total of 29 nights which are found to be variable based on all the above mentioned criteria i.e., C$\geq$ 2.57,  F $\geq$ 5 for more than 99$\%$ confidence level and $A_{var} \geq$ 0.05 mag.   Figure 1 shows the light curves for these INV  nights where time in Modified Julian Date (MJD) is plotted along X-axis and the brightness magnitude in R-band along Y-axis. Lower curve is the differential $LC$ for the two comparisons (5 and 6), to check the stability of that particular night, thus providing extent of uncertainty in the source values. The $rms$ values of these differential lightcurves for comparison stars are a measure of accuracy in our magnitude measurements. Upper curve (solid circles) shows the  calibrated  brightness magnitudes for the source. The plotted photometric errors  are of the order of a few milli-magnitudes.

\begin{figure*}

\includegraphics[width = 0.35\textwidth]{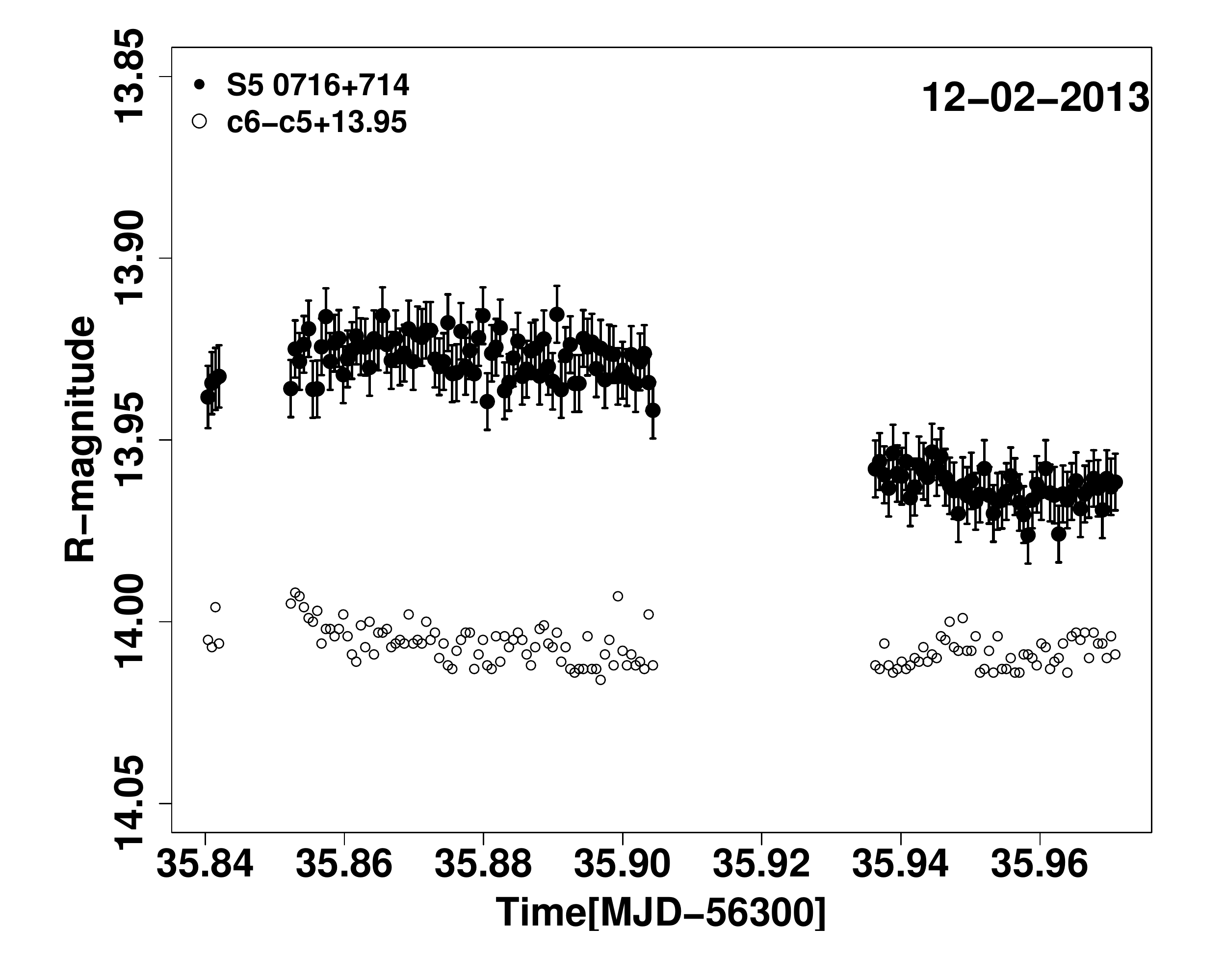} 
\includegraphics[width = 0.35\textwidth]{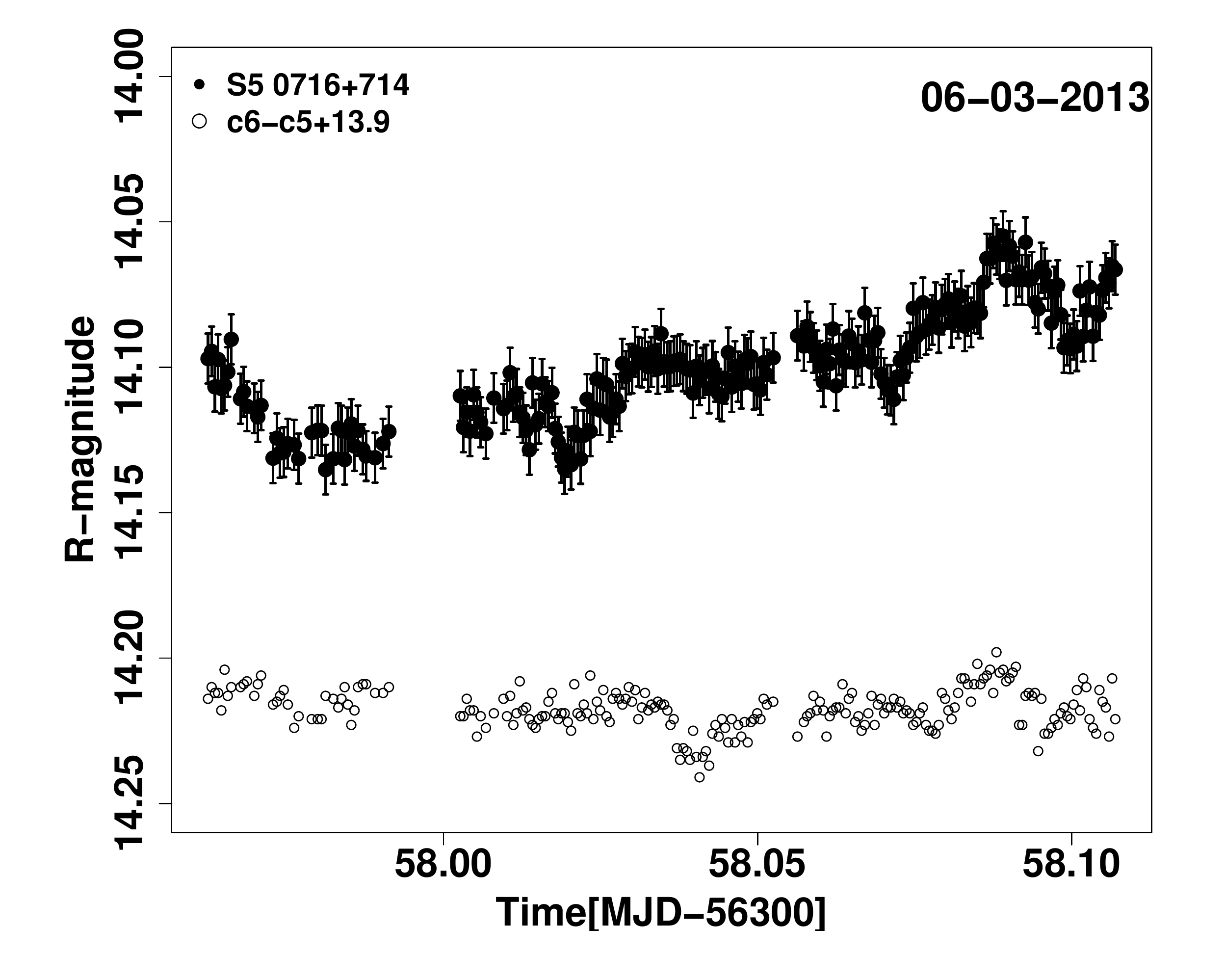}
\includegraphics[width = 0.35\textwidth]{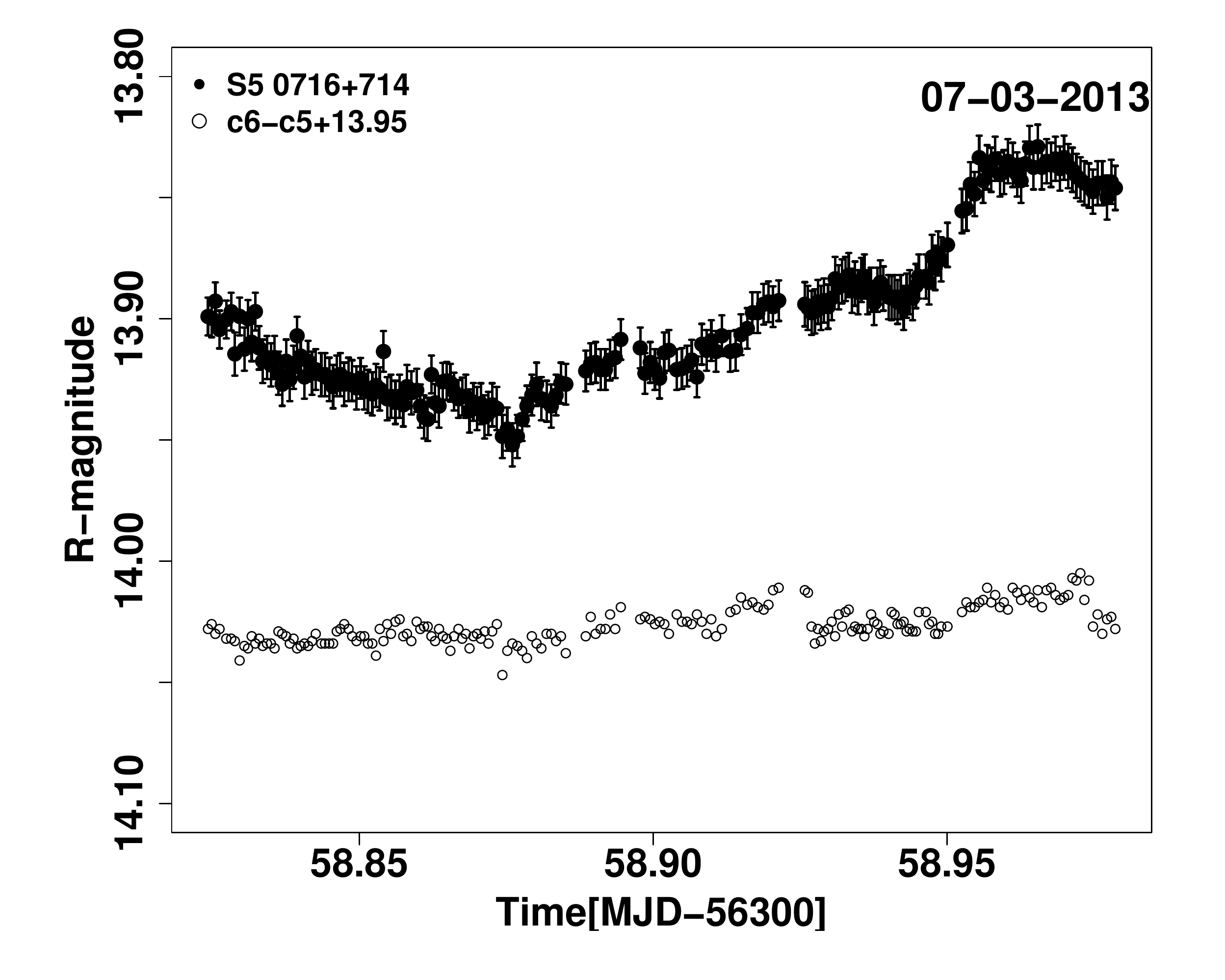}
\includegraphics[width = 0.35\textwidth]{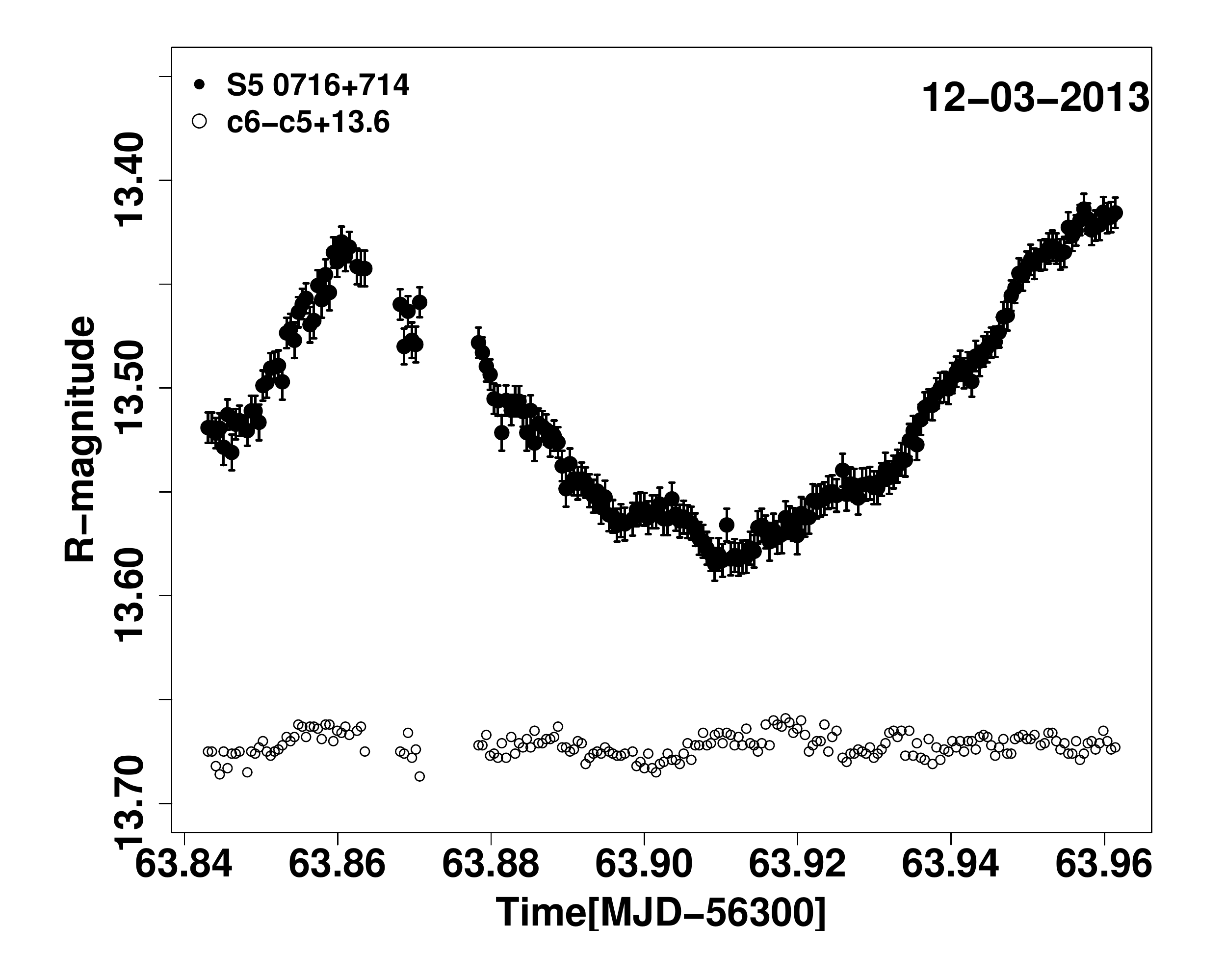}
\includegraphics[width = 0.35\textwidth]{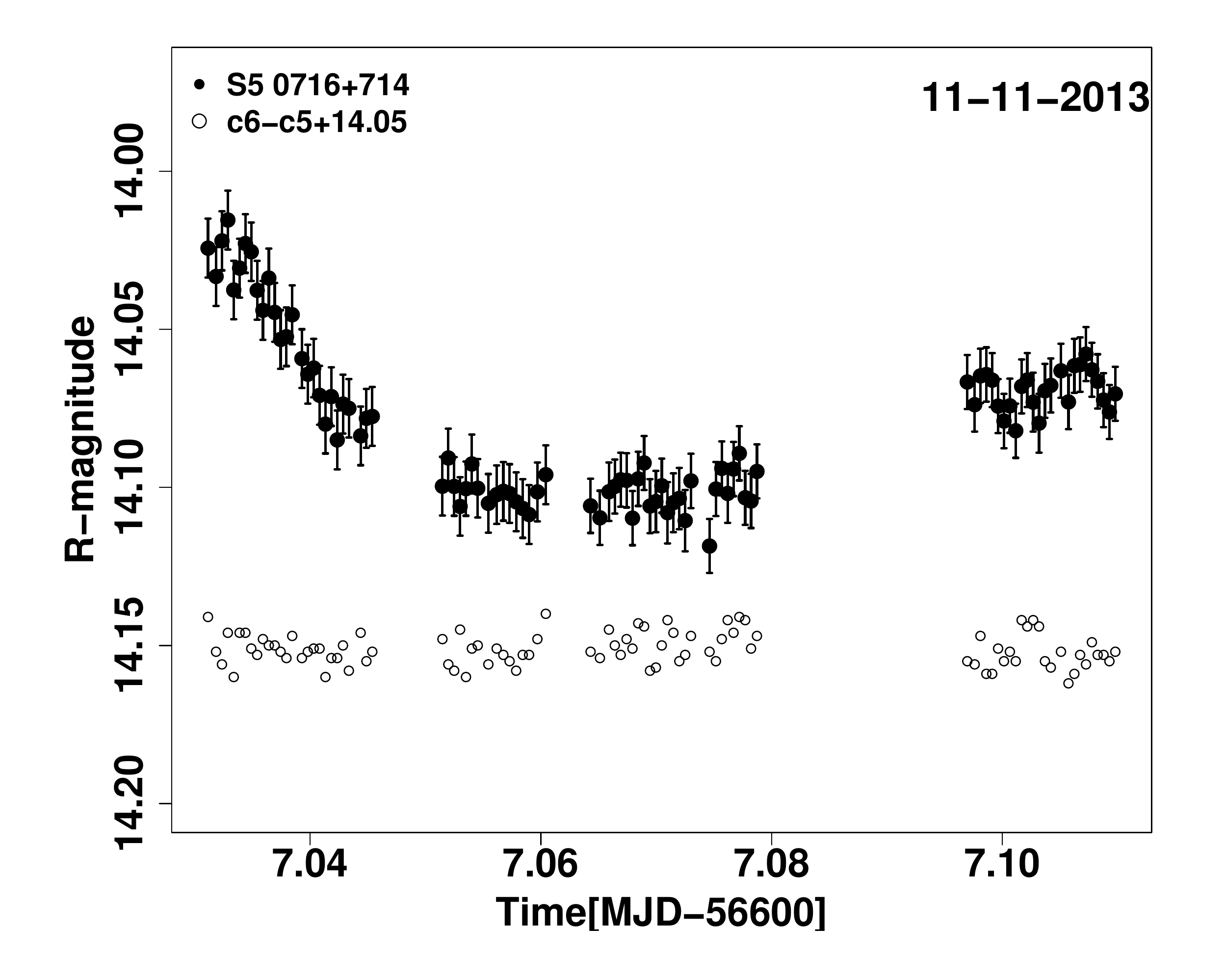} 
\includegraphics[width = 0.35\textwidth]{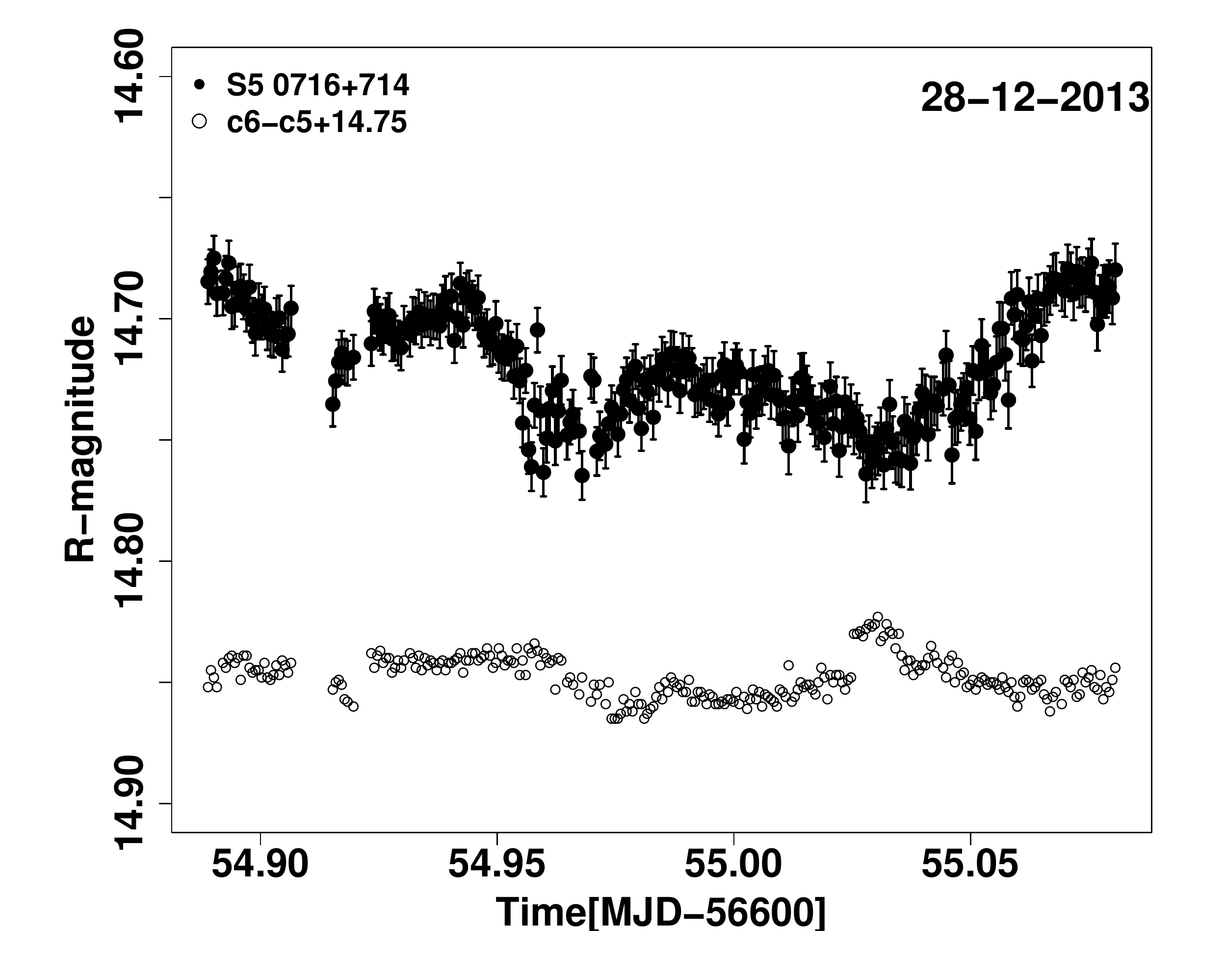} 
\includegraphics[width = 0.35\textwidth]{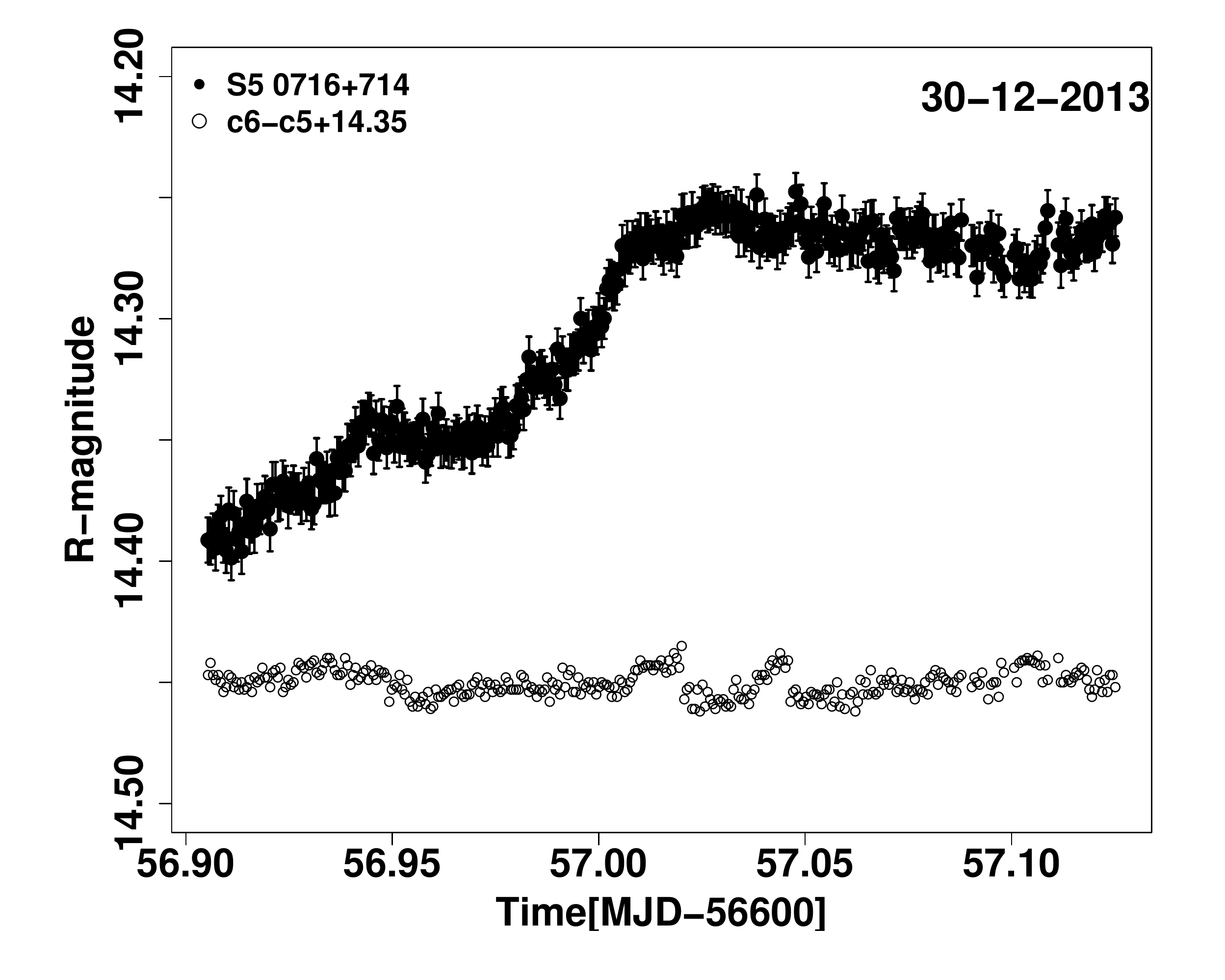}
\includegraphics[width = 0.35\textwidth]{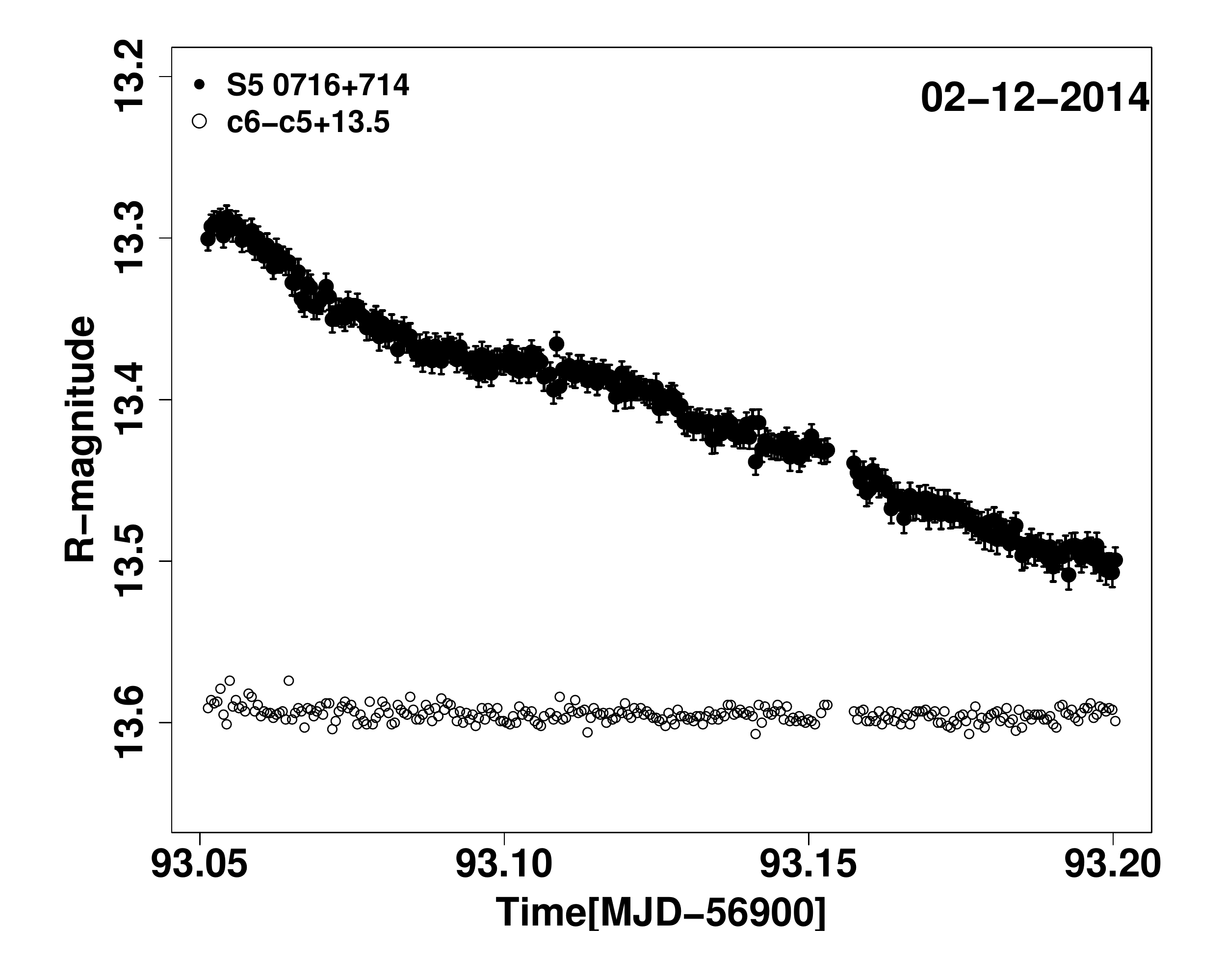}
\includegraphics[width = 0.35\textwidth]{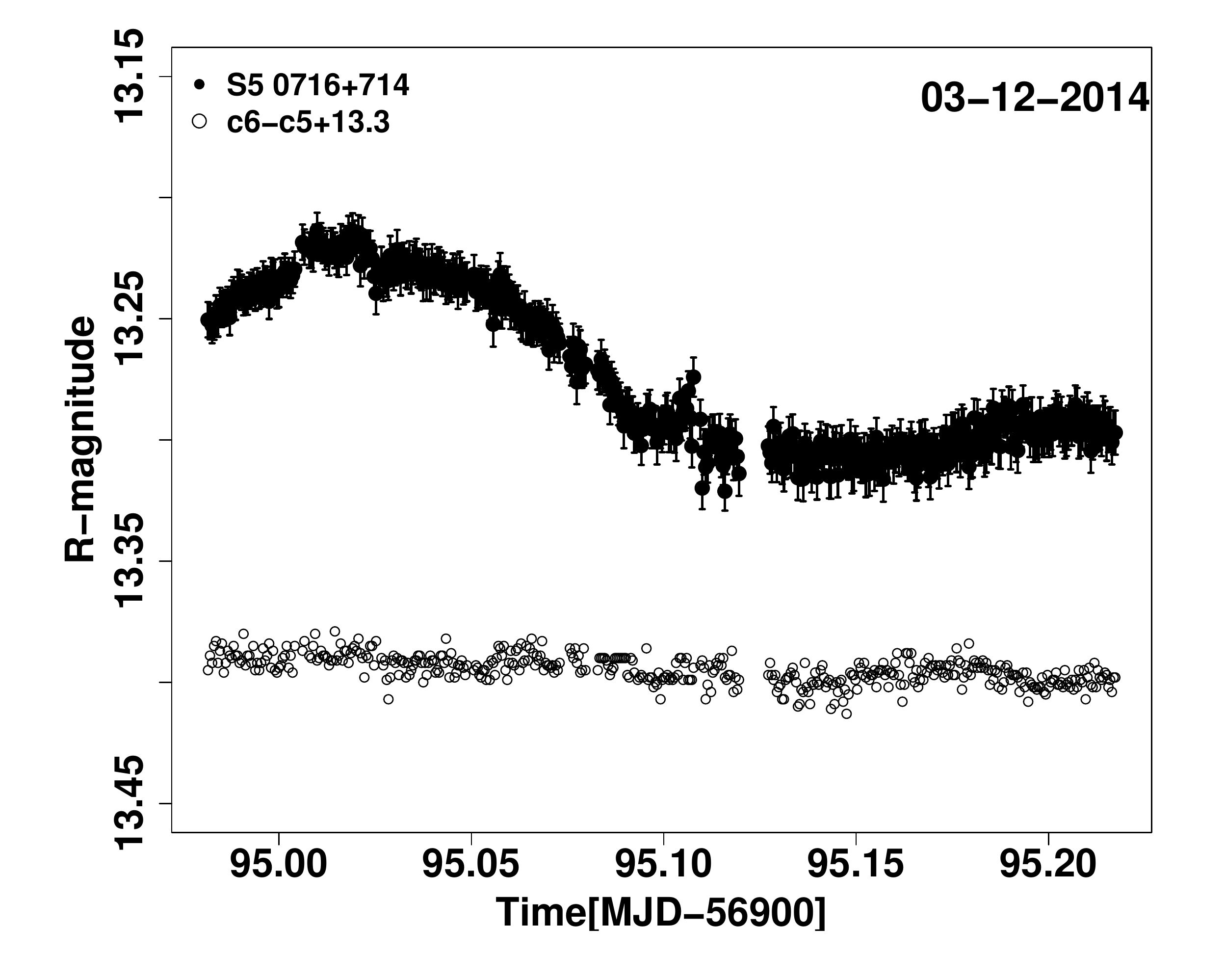} 

\caption{Intra-night light curves for the source S5 0716+714 on various nights during January, 2013 to June, 2015.}
\label{idv}
\end{figure*}

 The INV light-curves  (Figure \ref{idv}) feature monotonic  rise or fall,  slow rise or decay with  rapid fluctuations superimposed on them, alongwith a few $LC$s indicating to a possibility of  quasi-periodic oscillations with short timescale. It can be noted that the shapes of  most of the nightly lightcurves are different, as also reported by several other authors \citep[][ and references there-in]{chandra2011, kaur-3c2017, Hong2018} indicating that the emission processes are stochastic and complex in nature. A symmetric flare in a lightcurve would mean the cooling timescale is much shorter than the light-crossing timescale. On the night of 2013 February 12 (Fig. \ref{idv}), the brightness decays slowly with no distinct peak, with total change in the amplitude of variation by about 7.5\%. In the same figure, a slow increase in flux by about 0.07 mag in about 2.6 hr, with several rapid fluctuations superimposed (including one with 0.04 mag  in about 30 min) is noticed on 2013 March 6.  Next day $LC$  starts with slight decreasing trend but begins brightening up at MJD 56358.88, with a rapid increase after 1.44 hr leading to 0.06 mag (\textgreater  2 $\sigma$). The flux decreases up to MJD 56607.0 and then remains stable within errors on 2013 November 11. The INV $LC$ on 2013 March 12 shows interesting features with a brightening by 0.11 mag in about 30 min, followed by a decay of about 0.17 mag in about one hour. It starts increasing again reaching initial level of about 13.41 mag.   A slow decrease in flux and then relatively faster increase by about 0.07 mag within about 70 min characterizes the $LC $ on 2013 December 28(Fig. \ref{idv}). A significant increase in flux by 0.13 mag within about 2.9 hr is noticed on 2013 December 30, while on 2014 December 02 night, brightness decreases continuously, with no peak. On 2014 December 3, flux rises by 0.08 mag within 2.4 hr during the total monitoring time of about 6 hrs. 
 
 However, it is difficult to determine variability time scales accurately only from visual inspection of $LC$s  and therefore, in the next section we introduce and use structure functions and later analyze them to estimate required parameters.

\subparagraph{\bf Duty Cycle of variation:}
\smallskip
Most of the blazars show very high probability of variation even on intra-night time scales with an amplitude of variation of a few tenths of magnitude,  for example, CTA 102: \citet{Bachev2017}, 3C 66A: \citet[][ and references there-in]{kaur-3c2017}, S5 0716+714: \citet{chandra2011}. In order to quantify the probability of variation in a source, duty cycle of variation (DCV) is often used. The DCV is defined as the fraction of  total number of nights the source is monitored for,  which are found variable \citep{Romero1999}.  An expression to estimate DCV  is given by,

\begin{equation}
DCV = 100 \frac{\Sigma_{i=1}^n (N_i / \Delta t_i)}{\Sigma_{i=1}^n (1 / \Delta t_i)}  \%
\end{equation}
where, $\Delta t_i$ = $\Delta t_{i,obs} (1+z)^{-1} $ is duration of monitoring in rest frame of the source, 
and $N_i$ is $0$ or $1$ depending on whether the source is non-variable or variable, respectively. 

\smallskip
Several authors \citep[][ and references there-in]{wagner1995, chandra2011, dai2015} have reported INV $DCV$ for  S5 0716+714 ranging from 40\% to 100\% during their observations, which indicates that the source is almost always active. In our case, 9-nights out of a total of 29 nights monitored for more than two hours, are detected as confirmed variable ones. Thus based on  our observations during 2013-2015,  we get a value of  31\% as duty cycle,  which is on the lower side. Reasons could be that we monitored the source, by chance,  when it did not show much activity or our  duration of monitoring may not be sufficient.  \citet{Hong2018} monitored the source for less than one hour and reported a DCV of 19.57\% and, in another study done over 13 nights during 2012 January-February,  a value of 44\% \citep{Hong2017} was estimated, when the source was monitored for about 5-hours. In order to check for any connection between the INV shown by the source and the duration over which it was monitored, we calculated the duty cycle  with more than one hour  and two hour monitoring period. 

Out of the total 46 nights of observation during 2013-2015, we find 35 nights and 29 nights monitored for a minimum of  one hour and two hours, respectively. Based on these, we obtained  INV duty cycle values for the S5 0716+714 as 26$\%$ and 31$\%$, respectively, in two cases. It, therefore,  indicates that longer the duration of monitoring, higher will be the probability of finding a source variable, i.e., a higher DCV. 

\noindent
\subparagraph{\bf Rise \&  fall rates of variation in INV lightcurves:} 
\smallskip
To investigate the extent of the  intra-night variability  of the source, we determined rate of change in magnitude (rise/fall) on each INV night for S5 0716+714 by fitting a line segment to light curves.  These rates of variation are given in  Table \ref{t_risefall}.

\begin{table*}
%\label{t_risefall}
\centering
\caption{Details of the rates of rise/fall in the magnitude for  INV nights. $\Delta{m}_{+}$,  $\Delta{m}_{-}$ represent the source brightening or  dimming, respectively. }
\begin{tabular}{cllrc}
\hline
\hline
Date &Trend &Rise/Fall mag & Duration  & Rate=$\Delta$m/$\Delta$t\\
& &($\Delta{m}_{+}$/$\Delta{m}_{-}$) &(in minutes) &(mag$/$hr) \\
\hline
12-02-2013 &Fall&0.05 ($\Delta{m}_{-}$)& $~$ 198 	&0.015 \\
06-03-2013&Flickering over&0.08 ($\Delta{m}_{+}$)&$>$150  & 0.02\\
&a monotonic rise& & &\\
07-03-2013	&Fall  		& 0.05	($\Delta{m}_{-}$)		&$~$ 72 	&0.03\\
			&Rise &	0.12 ($\Delta{m}_{+}$)		& $~$144  	&\\
12-03-2013	&Sine like 	&0.10 ($\Delta{m}_{+}$)			& 30  			& 0.05\\
			&pattern			&0.20 ($\Delta{m}_{-}$)			& $~$72 	&\\
			&				&0.20 ($\Delta{m}_{+}$)			& $~$72 	&\\	
11-11-2013	&Fall 	&0.08 ($\Delta{m}_{-}$)		& $>$ 72 		&0.04\\ 
%			&Rise		&14.06 mag	(stable)		&at MJD=56607.10 		&\\
28-12-2013	&Fall with		& 0.08 ($\Delta{m}_{-}$)		&72  			&0.38\\
			&flickering		& 0.02	($\Delta{m}_{-}$)			& $>$ 20 &\\
30-12-2013	&Monotonic rise		&0.14 ($\Delta{m}_{+}$)			& 144  	&0.07	\\ 
			
02-12-2014	&Monotonic fall	&0.02 ($\Delta{m}_{-}$)		&216 		&0.05\\
03-12-2014	&Sine like		&0.08 ($\Delta{m}_{-}$)		&288 		&0.02\\
\hline
\label{t_risefall}		
\end{tabular} 
\end{table*}

                                                                                                                                                                                                                                                                                                                                                                                                                                                                                                                                                                                                                                                                                                                                                                                                                                                                                                                                                                                                                                                                                                                                                                                                                                                                                                                                                                                                                                                                                                                                                                                                                                                                                                                                                                                                                                                                                                                                                                                                                                                                                                                                                                                                                                                                                                                                                                                                                                                                                                                                                                                                                                           During our observations, 2013 February 12 and 2013 December 28 represent the nights with minimum and maximum rates of  change in the  magnitudes of the  source  with 0.015 mag hr$^{-1}$ and 0.381 mag hr$^{-1}$  (cf,Table \ref{t_risefall}), respectively. The rate of brightness change on 2013 December 28 happens to be one of the fastest for this source. Earlier,  \citet{chandra2011, man2016} have reported 0.38 mag/hr  \& 0.35 mag/hr rates, respectively. Rate of change in the brightness magnitude as high as 0.43/hr has been reported for the  PKS 2155-304 \citep{Sandrinelli2014}. The source showed smooth decline in its brightness  by 0.05 mag on February 12 with 7.60 \% amplitude of variability. On 2013 March 6, S5 0716+71 became brighter by 0.08 mag in 3 hours with rapid fluctuations (few tens of minutes duration) superposed over the day-long trend showing 7.58$\%$ amplitude of variation in the light curve. On 2013 March 7, brightness decreases from 13.90 mag to almost 13.95 mag within an hour, after which source brightened by more than 0.1 mag in next 3 hours with a rate of change of 0.03 mag hr$^{-1}$ as mentioned in the Table \ref{t_risefall}. On 2013 March 12, light-curve shows a sine like feature, with rising (0.1 mag in 30 min) - declining (0.2 mag in $~$72 mins) - rising ($>$0.2 in about 72 min approx.) trends in brightness over the duration of more than three hours.

 The light curve on 2013 December 28  showed sharp 
rise/fall magnitudes over two peaks and again showed a rising trend with overall change in magnitude by 0.38 mag hr$^{-1}$  (see Table 1).  However, the features in the light curves are asymmetric in nature, which rules out variation being caused by extrinsic/geometric mechanisms. The variability in blazars is stochastic in nature at almost all timescales. The flares, therefore, appear to be produced independently and any similarity or difference might reflect different scales of particle acceleration and energy dissipation \citep{Nalewajko2015}. The variations in blazars are caused largely in the jet but it is difficult to ascertain whether these are intrinsic or geometric in nature. Intrinsic variations are dissipative and  irreversible in time. Hence they should cause asymmetric flares. The geometric variations, on the other hand, are symmetric in time \citep{bachev2012} and achromatic in nature. The intrinsic variability could be due to fast injection of relativistic electrons and radiative cooling and/or escape of the particles or radiation from the emission zone.  The symmetric flares, however, might result if cooling time scale is much shorter than the light crossing time \citep{chatterjee2012}.

 The  INV $LC$s are, in general,  asymmetric and complex indicating the random/turbulent nature of the flow inside the jet. Based on the visual inspection of these curves, we identify three observed trends:
 
a) Rapid intra-night changes in the source flux, indicating to the violent, evolving nature of the shock formed in the jet. It might be either  due to the presence of oblique shocks or instabilities in the jet.
\smallskip
b) The steady rise or fall in the light-curve during a night indicates to the light crossing timescale to be shorter than the cooling time scale of the shocked region. It is when data series is shorter than characteristic time scale of variability. The cooling times shorter than light crossing time would have resulted in symmetric light-curves \citep{Chiaberg_Ghis1999, chatterjee2012}.
\smallskip
c) The small-amplitude rapid fluctuations (asymmetric in shape) superimposed over slowly varying  light curve suggest small scale perturbations in the shock front or  oscillations in the hot-spots downstream the jet and may not be associated with size of the emission regions.

\begin{table*}
%\label{t_idv}
\textwidth=7.0in
\textheight=10.0in
\vspace*{0.5in}
\noindent
\caption{Details of the INV nights for  the  source S5 0716+714 during 2013-2015. Column 1 to 11 present; date, MJD, observation start time, duration, no. of images, average magnitude with error, test parameter C, amplitude of variation, variability time scale, SF parameters;  k \& $\beta$. }
\begin{tabular}{ccccccrrrcc}
\hline
\hline 
Date of 	   &MJD & $T_{start}$ &  Duration &  N\footnote{Number of data points}    &  $\bar{m}$ $\pm$ $\sigma$ & C  &  $A_{var}$  & $t_{var}$ &k &$\beta$\\
 observation	& 	& (hh:mm:ss)   & (hrs)	 &   &  & 	& ($\%$)& 	\\		
\hline

 12-02-2013   & 56335.84038 & 20:10:09    & 3.28  & 195 & 13.94 $\pm$ 0.02 & 2.63   & 7.60  & \textgreater 2.36 hr &0.31 &4.82\\
%    &   &    &   &  &   &    &  \\
%
 06-03-2013  & 56357.96250 & 23:06:00    & 3.46  & 237 & 14.10 $\pm$ 0.02 & 2.72   & 7.58& \textgreater 1.68 hr &0.99 &1.67\\
 %        &   &    &   &  &   &    &  \\
%
 07-03-2013 	 & 56358.82424	& 19:46:54     & 3.71  & 203 &  13.90 $\pm$ 0.03	&  4.46 & 11.38&  2.04 hr &1.71 &1.81\\
 %&   &    &   &  &    &    &  \\
%
 12-03-2013	& 56363.84308  & 20:14:02	  & 2.83 		& 229 & 13.50 $\pm$ 0.05		& 8.90	& 15.61 &  
1.11 hr &0.08 &1.21 \\
% &   &    &   &  &    &    &  \\
%
 11-11-2013		&  56607.03115	&  00:44:51	&  1.88 	& 139  &  14.08 $\pm$ 0.02	&  4.62	&  9.75 &  
0.96 hr  &3.61 &1.33\\
%&   &    &   &  &    &    &  \\
%
 28-12-2013	& 56654.88889	& 21:20:00	 & 2.48 	& 284 & 14.72 $\pm$ 0.02		& 2.65	& 11.89 &  
0.76 hr  &1.14 &0.97 \\
%&   &    &   &  &    &    &  \\
%
 30-12-2013		& 56656.90536 & 21:43:43	 & 2.26	 & 350 & 14.30 $\pm$0.05			& 7.55	& 15.38 &  3.1 hr &0.88 &1.31\\
%&   &    &   &  &    &    &  \\
%
 02-12-2014		& 56993.05133  & 01:13:05	 & 4.58	& 284 & 13.41 $\pm$0.06		& 12.37	& 20.50 
&  3.54 hr 	&2.29 &2.34\\
%&   &    &   &  &    &    &  \\
%
 03-12-2014		& 56994.98155  & 01:13:55	 & 5.97	& 454 & 13.27 $\pm$0.03		& 5.23	& 10.07&  3.89 hr &2.32 &2.41	\\
&   &    &   &  &    &    &  &  & & \\
\hline	
\label{t_idv}		
\end{tabular}
\end{table*}

\subsubsection{\bf Variability timescale, size of emission region and black hole mass}
%{\bf The blazar continuum spectra show very weak or no emission lines, the fastest variability timescales are the diagnostic tool to study the compact regions, assuming the activities are happening in vicinity of the blackhole. The characteristic timescale is very crucial to constrain the size of the emission region and the mass of a blackhole\citep{miller1989,  Xie2002,liang2003,   dai2015}. By knowing the extent of the jet opening angle, the location of the compact emission regions could be roughly estimated \citep{ahnen2017, nav1es2017}.}

It is important to know the characteristic timescale of intra-night variability which can constrain the emission size and structures of the blazar emission zones.  If we consider the shape of the jet as conical close to its origin, the opening angle and the extent of vertical expansion of the jet can provide us a rough estimate of the location of emission region with respect to the supermassive black hole \citep{ahnen2017, nav1es2017}. 

The rapid variations with duration of a few hours originate, perhaps, in the close vicinity of the central engine where jets are launched and might be caused by a combination of accretion disk instability, shock propagating within the jet, and/or particle acceleration and consequent radiative cooling near the base of  jet \citep{ulrich1997}. 
This assumption is also used to estimate the mass of the black hole, which is difficult to determine otherwise as BL Lacs do not show emission lines.  Since the INV light curves are complex, we  use statistical tools, described here, to  discuss features in the intra-night light curves and estimate INV timescales and any possible quasi-periodicity.

\subparagraph{\bf Structure function:} 
\smallskip
 The structure function (SF) described by \citet{simonetti1985, gliozzi2001} provides   information about characteristic timescale ofvariability for flat- and steep spectrum radio sources by analyzing their light curves. In order to estimate the characteristic variability timescale, we used first order structure function for a magnitude data series, defined as,
\begin{equation}
\label{eq5}
SF(\tau_{i}) = \frac{\Sigma{[M(t + \tau_{i}) - M(t)]^{2}}}{N}
\end{equation}
where, M(t) is the magnitude at time $t$ and $\tau_{i}$ is the time lag. 
The chi-square method is used to fit the structure function where from minimum variability timescale and corresponding errors are estimated \citep{zhang2012}. The SF reveals extent of changes in the magnitude as a function of time between two observations. In this curve of growth of variability with time lag, a plateau (change of slope or saturation of SF) might indicate presence of a characteristic time.

\begin{equation}
  SF(\tau)= \left \{
    \begin{array}{c l}
    k{\tau}^{\beta},   & \tau\underline{<}\tau_o , \hfill \\
    C,   & \tau>\tau_o.
  \end{array}\right\}
\end{equation}

where, $\tau_{o}$ is characteristic timescale with 1$\sigma$ uncertainty and $\beta$ = $\frac{dlog(F)}{dlog(\tau)}$ is logarithmic slope in $\tau-SF$ plane characterizing the nature of the variability and physical  processes. If the value of $\beta$ is close to 0, it indicates flickering noise, while $\beta$ $\geq$ 1 indicates turbulent process in jet (or shot-noise) responsible for the changes.

\begin{figure*}
\includegraphics[width = 0.3\textwidth]{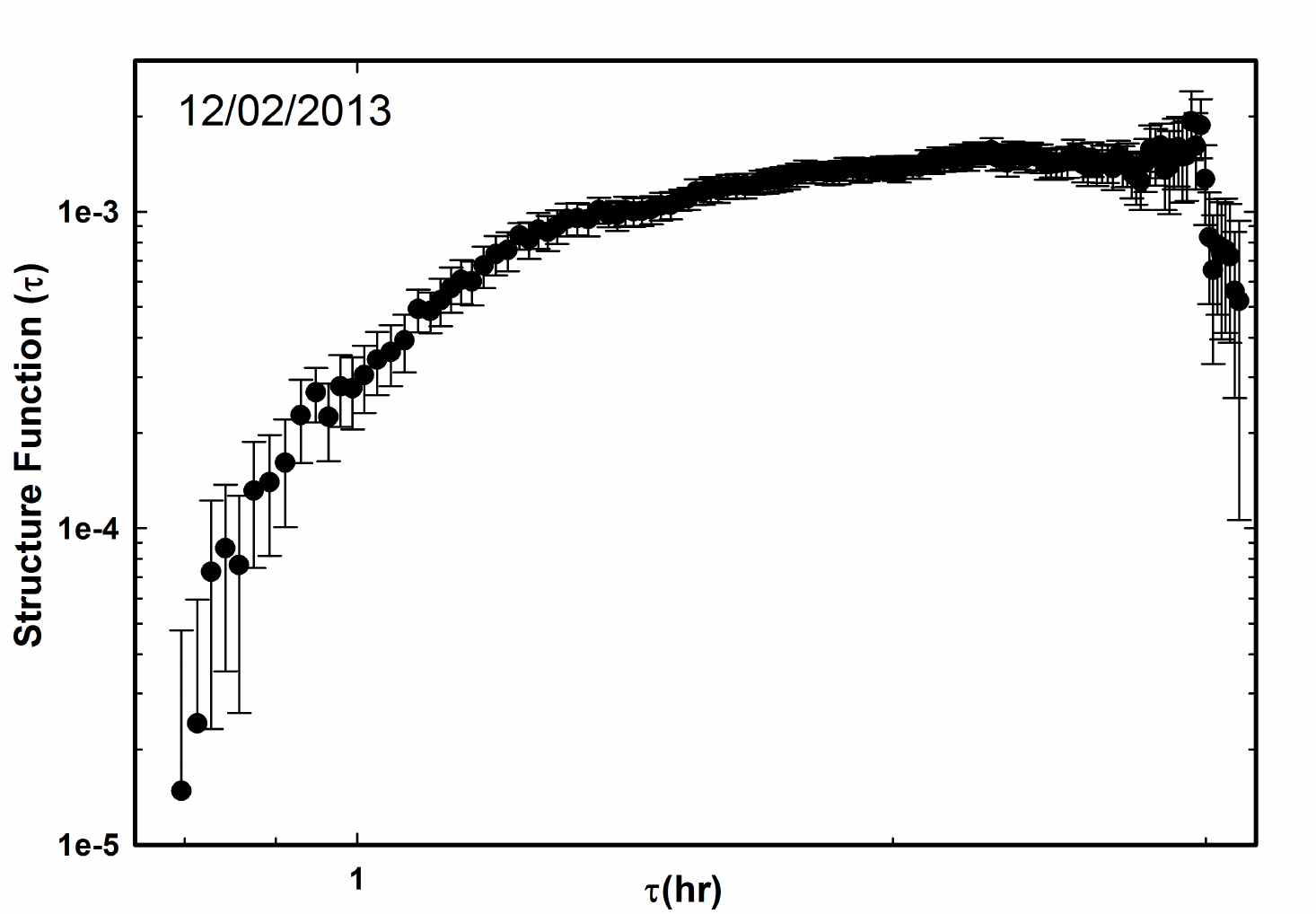} 
\includegraphics[width = 0.3\textwidth]{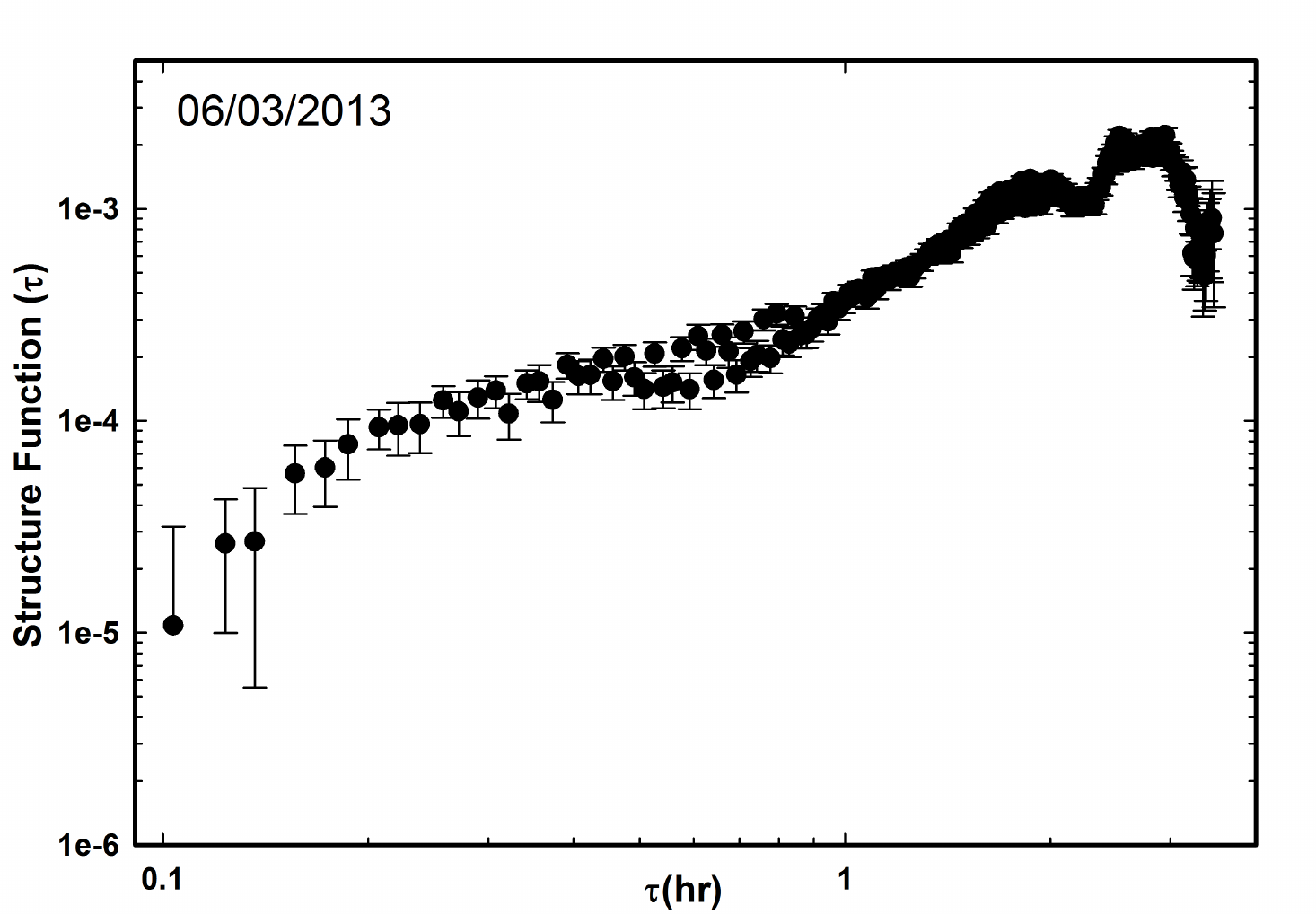} 
\includegraphics[width = 0.3\textwidth]{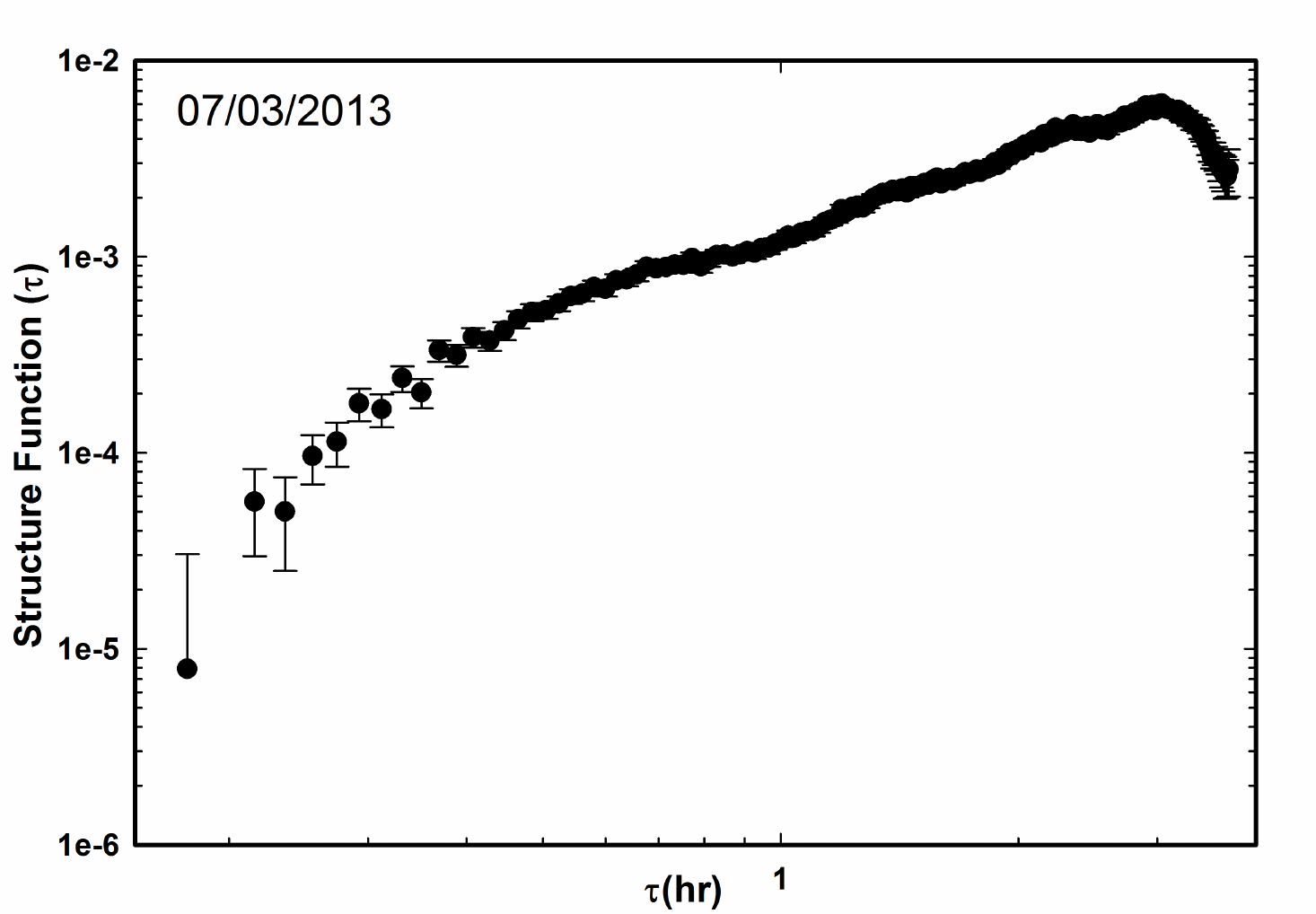}
\includegraphics[width = 0.3\textwidth]{sf060313.PDF}
\includegraphics[width = 0.3\textwidth]{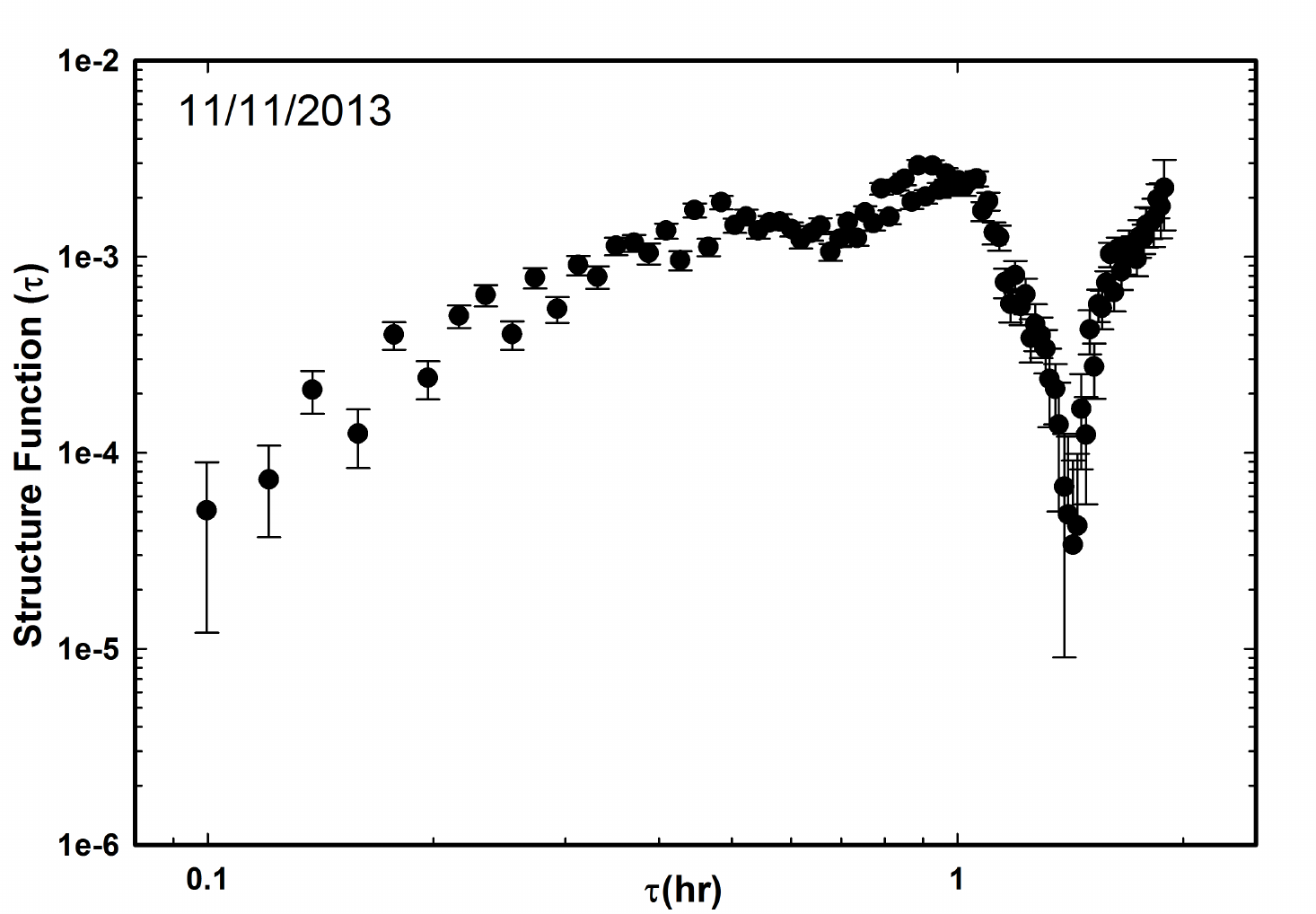}
\includegraphics[width = 0.3\textwidth]{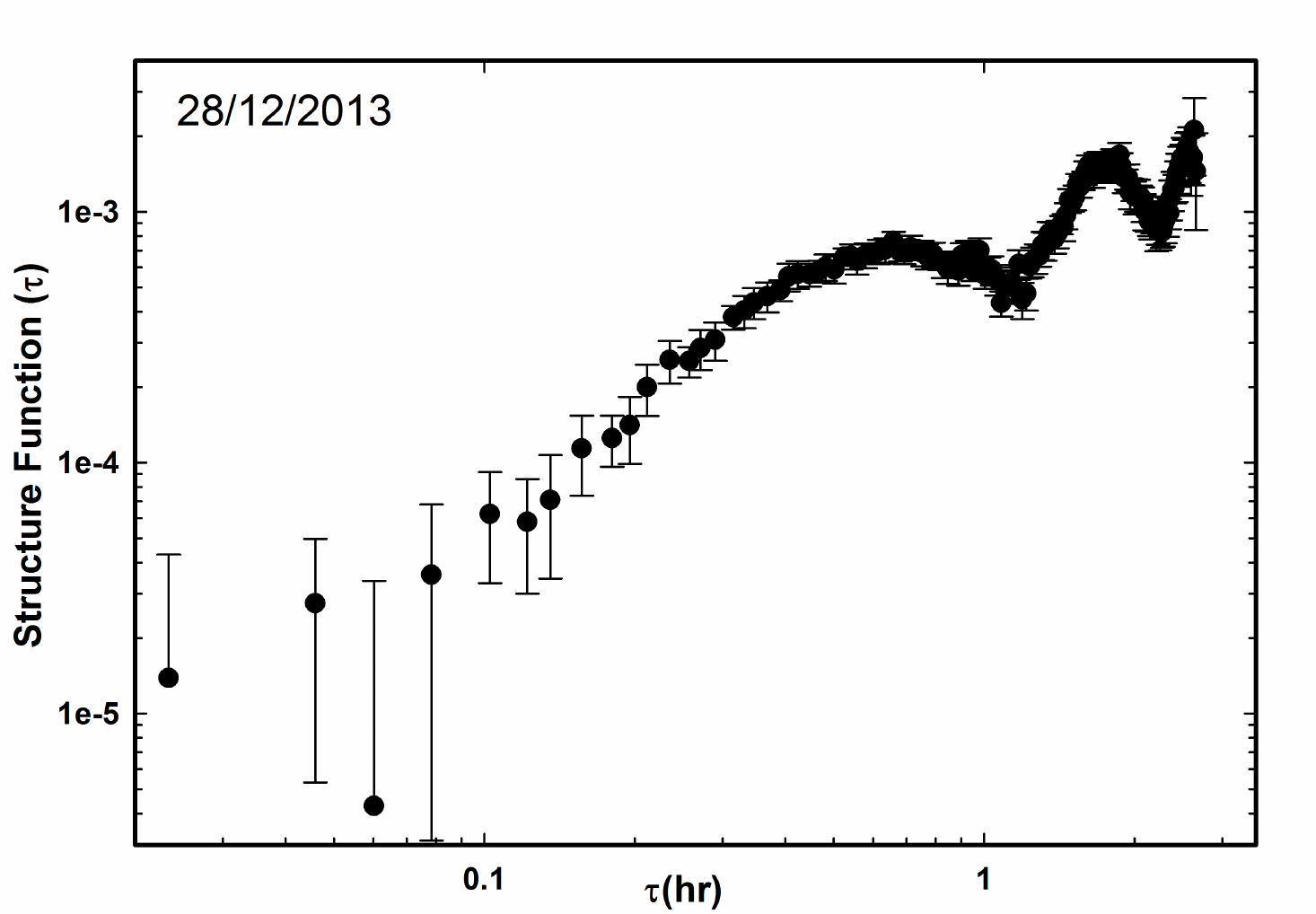}
\includegraphics[width = 0.3\textwidth]{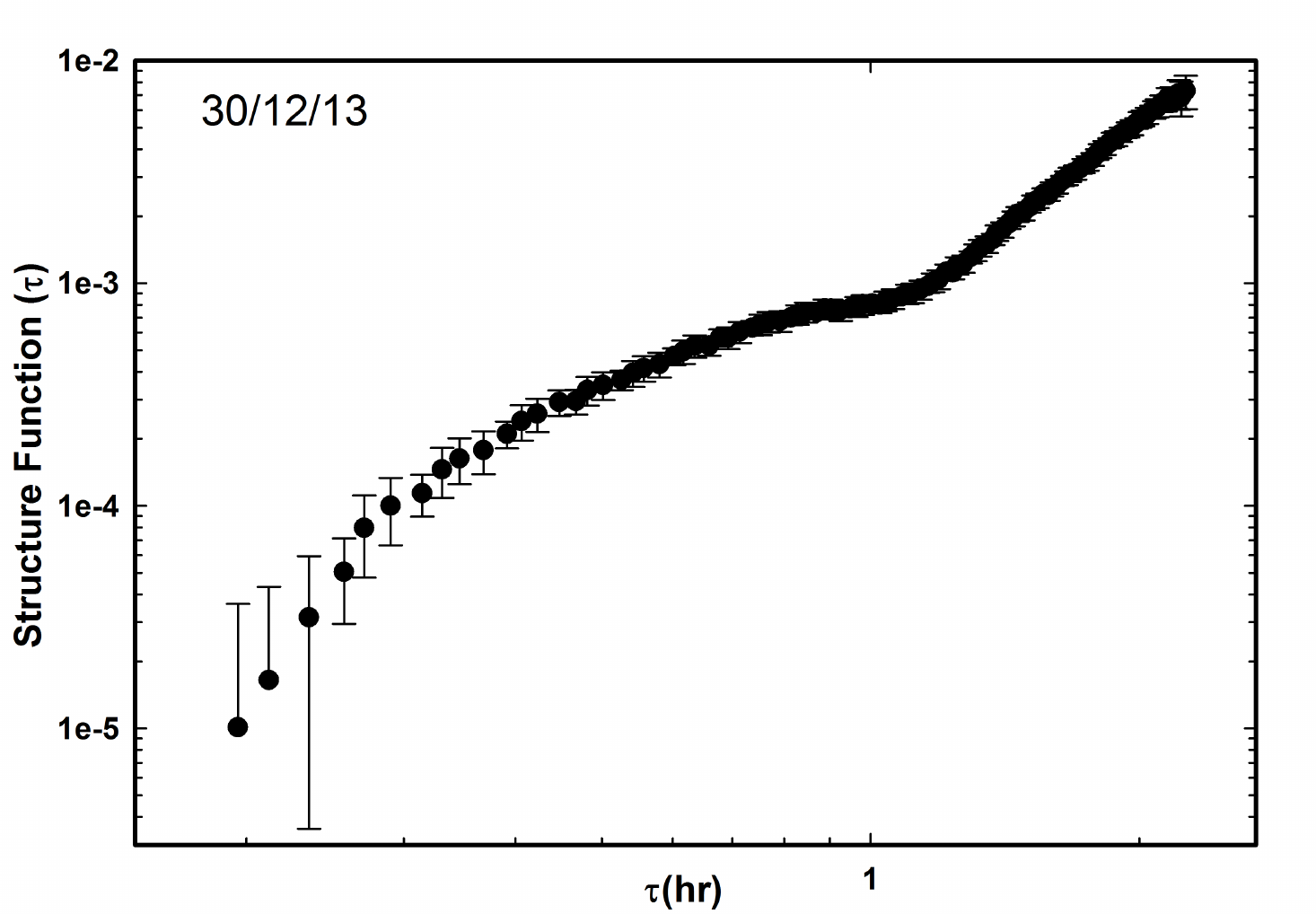} 
\includegraphics[width = 0.3\textwidth]{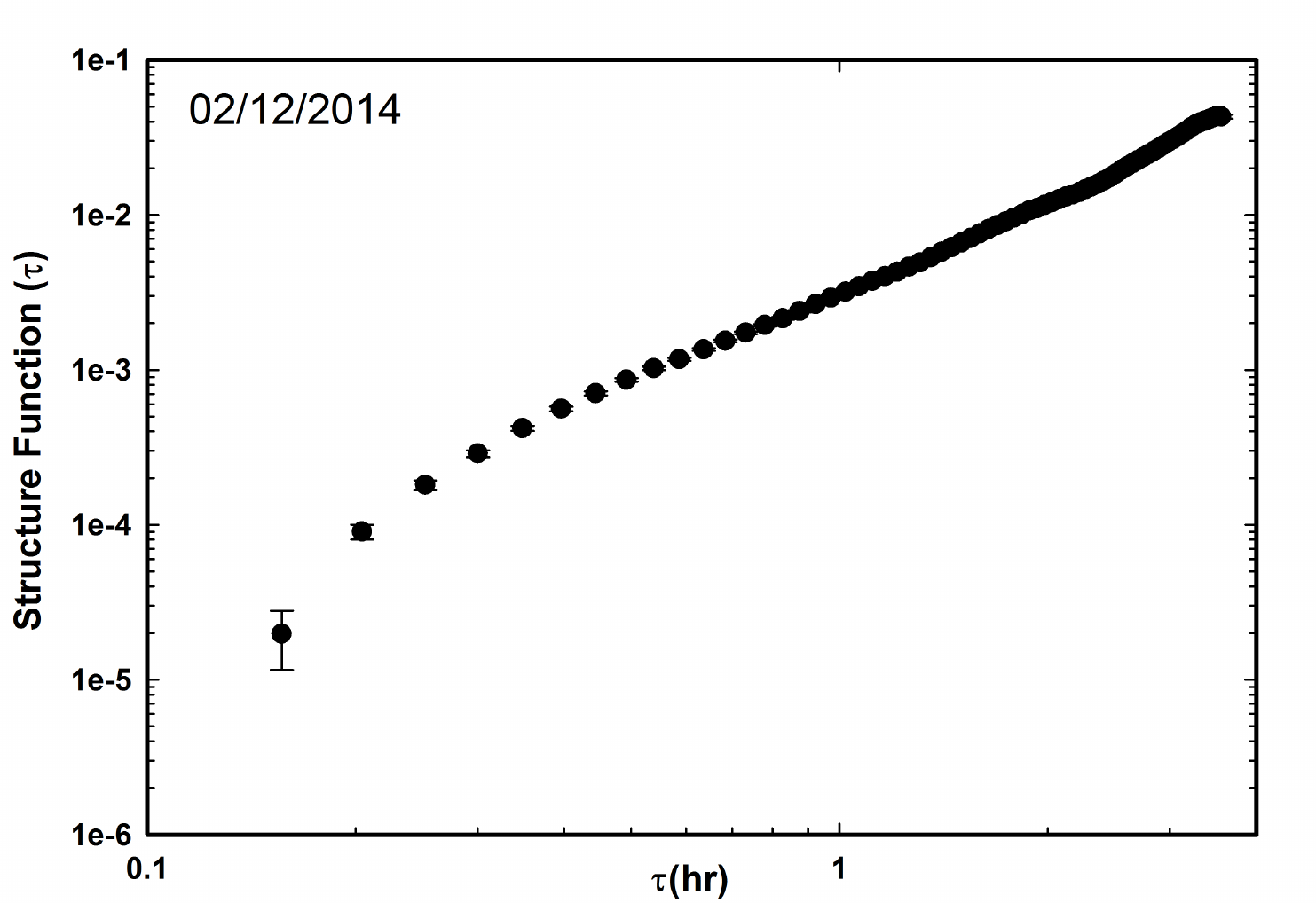}

\caption{Structure functions for the INV nights are plotted for the source S5 0716+714 during 2013-2015. X-axis represents time lag in hours.}\label{sf}
\end{figure*}

\smallskip  
Figure 2 shows the structure functions for  the INV nights. It is seen that first order SF for several nights does not show any plateau, that means that the characteristic timescale of variability is  longer than the length of the observational data \citep{dai2015}.  The local maximum following the smooth rise in the SF-$\tau$ plane reveals time-scale of variability introduced by the presence of   minimum and maximum or vice-versa in the curve. If the SF consists of more than one  plateau with  slopes ($\beta$) following a power-law trend, presence of multiple timescales is inferred.  If periodicity is present, it will be seen as local minima in SF after the occurence of a  local maximum. The difference between two minima gives the time period. 

On several nights (2014-12-02, 2013-03-07, 2013-11-11,  \& 2013-02-13), SF shows continuous increase with no or a feeble plateau, indicating that characteristic time scale of variability is longer than the monitoring period, giving only a lower limit of the variability time scale. While $LC$ for the night 2013 March 12 shows several features, SF shows only one plateau and then a dip at about 2.2 hr. On December 13, 2013, SF shows a discernible peak with a time scale of 3.11 hr, followed by a rise. SF for INV night of 2014 December 14 shows a plateau at 2.89 hr. 
As can be seen, INV night 2013 December 28 shows a plateau in its SF giving the shortest characteristic timescale of variability of about 45.6 minutes during our observing  campaign. 

 The quantitative values of various parameters related to INV nights, including SF parameters,  are given in Table 2.  In this table,  the column 1 and 2 represent the date of observation (in dd-mm-yyyy and MJD format, respectively), column 3 represents the start time of the observations, the duration of monitoring and the number of data points (images) are given in column 4 and 5, column 6 presents the average magnitude and the associated errors in the source brightness, column 7 and 8 contain the values of the  statistical parameters, i.e., C and $A_{var}$, respectively.  The variability timescales for INV nights are shown in column 9 along with the structure function parameters, normalization constant(k) and logarithmic slope($\beta$) in columns 10 $\&$ 11, respectively.
 
The INV lightcurves feature several complete events with specific time scales. Applying light travel time arguments, these time scales can be used to put  limits on the size of the emission regions responsible for the variation in flux. The shortest characteristic timescale puts constraint on the size of emission region \citep{elliot1974}. Using the characteristic timescales obtained from light curve and SF analysis, the size of the emission region,

\begin{equation}
\label{R}
R \leq \frac{\delta c \Delta{t_{var}}}{(1+z)}
\end{equation}

where, $c$ is speed of light, $\Delta{t_{var}}$ is minimum timescale of variability, $\delta$ is the Doppler factor and $z$ is the redshift of the source (z$=$0.32).  When considering long-term behaviour, various authors have used different values of the Doppler factor, $\delta$.  \citet{Bach2005} use Doppler factors 13 to 25 when viewing angle changes from 5$^\circ$ to 0$^\circ$.5. \citet{nesci2005}  adapted  a value of 20, while \citet{fuhrman2008}  apply a range 5 to 15 for the Doppler factor. We have used a value of 15 in this work, taking into account  the brightness of the source during the observed period. Thus, using $\Delta{t_{var}}$= 45.6 minutes as the characteristic time scale of variability, the estimated size of the emission region  is of the order of $\approx$ $10^{15}$cm. Apart from this shortest time scale of variability, other time scales estimated on  other nights indicate different sizes of emission regions in the jet. The longest time scale of variability detected in the present study is 3.89 hr which corresponds to a size of  4.8 $\times 10^{15}$ cm in the source frame. All these emitting regions are very compact and close to the black hole, within the BLR region.

Mass is one of the most important properties of a black hole. There are two categories of methods to determine the mass of a black hole in AGN; primary and secondary.  While there are direct, primary  methods applicable to nearby black hole systems  where motion of the surrounding  stars and gas under the influence of black hole, are traceable \citep{vestergaard2004}, it is very difficult to have an estimate of their masses at high redshifts. In the secondary methods, mass of the black hole is estimated by resorting to  approximations, e.g., using a parameter with which black hole mass is correlated. There are several methods which fall into this category. However, for the sources  which  do not show any emission line and whose host galaxy is also weak/non-detectable, which is the case for BL Lac type sources, it becomes extremely difficult to estimate the mass of black hole. For such systems, the variability time scale can provide a rough estimate of the black hole mass,  assuming that the shortest time scale of variation is governed by the orbital period of the inner most stable orbit around a Kerr (maximally rotating) black hole. \citet{miller1989} claim the origin of microvariability to arise from a location very close to the central engine based on the fast variability time scales, while \citet{marscher1992} associate their location somewhere down the jet and perhaps near the sub-mm core , caused by turbulence.  

Many authors have estimated masses of black holes residing  in the BL Lac sources following the earlier \citep{miller1989} approach using shortest variability time scales \citep{Fan2005,gupta2008, Rani2010, chandra2011, kaur-3c2017,  Xie2002, liang2003,   dai2015}. 
Here we use this method to estimate the mass of  a Kerr black hole at the center of S5 0716+71 using the expression \citep{abram1982, Xie2002}

\begin{equation}
M =1.62 \times {10}^4 \frac{\delta \Delta t_{min}}{1+z} {M}_{\odot},
\end{equation}

where,  $c$ is the speed of light, z, the redshift and $\delta$ is the Doppler factor. Taking the shortest variability time scale, $\Delta{t_{min}}$= 45.6 min and Doppler factor as 15,  we estimate the  mass of the Kerr black hole to be 5.6 $\times 10^{8}\, M_{\odot}$,  which is in close agreement with other values including a value of $1.25\times 10^8 M_{\odot}$  \citep{liang2003} obtained by using optical luminosity. \citet{bhatta2016} linked plateau in the $LC$ to the characteristic timescale for developing outflow within the jet base, equivalent to the innermost stable orbit and obtained the value of black hole mass as $4\times 10^9$ (maximally spinning BH) and   $3\times 10^8\, M_\odot$ (lowest spin BH). \citet{agarwal2016} obtained a value of 2.42$\times10^9 M_{\odot}$ for the black hole mass in S5 0716+714. \citet{Hong2018} estimated mass of the black hole as $5\times10^6\, M_{\odot}$ using 50 min QPV originated from the inner most orbit of the accretion disk.
\smallskip
Using  the black hole mass, M$_{BH}$, estimated here, the Eddington luminosity can be estimated from the following expression given by \citet{witta1985},
\begin{equation}
L_{Edd} = 1.3\times 10^{38} (M/M_{\odot})  erg\, s^{-1}
\end{equation}
 which, in case of the S5 0716+71 comes out to be about  7.28$\times 10^{46} erg\, s^{-1}$. 

\subparagraph{\bf  Quasi-periodic variability:} 
\smallskip
Another very interesting albeit highly debatable issue is the possible presence of periodicity in the blazar light curves. Claims for their existence have been made in optical bands \citep{lainela1999, fan2000}. A few INV $LC$s indicate  the presence of  possible quasi-periodic variations, also noticed in this source by \citet{wu2005, gupta2008, Rani2010, poon2009, man2016}, with periods varying from 15 minutes to 1.8 hr. The presence of such features, if genuine,  can be explained by the light-house effect \citep{Camenzind1992}, plasma moving in a helical magnetic field or micro-lensing effect etc. Recently \citet{Hong2018} obtained 50 min QPV from the observations during 2005 - 2012 when S5 0716+71 was in a relatively fainter state. They opined that the QPV is caused by the activity in the inner most orbit of the accretion disk.   In the present case, the variations seen on timescales of a few hours with asymmetric profiles rule out the possibility of the micro-lensing as the  mechanism.  Here, flares could be  caused by a sweeping beam whose direction changes with time due to helical motion. To estimate variability timescales and/or periodicity (if present) in our $LC$s, we use structure function and periodogram analysis.

The SF for the night of 2013-03-12 shows only one minima at about 2.2 hr while that for 2013-12-28 gives two minima at 1.25 hr and 2.3 hr, giving a possible period of about 1.2 hr.

 However, since these periods are either closer to the length of the data series and/or flux enhancements are less than 3$\sigma$, existence of periodicity is doubtful. Also, these features may not significantly represent departure from a pure red-noise power spectrum. Though the quality of the light curve presented here, in particular its dense sampling, is  good enough for a search of hour-long QPVs,  the fact that we did not find such QPV at a significantly high level to claim the detection, is  meaningful in itself. It implies that no persistent periodic signal exists in the source within the analyzed variability timescale domain.

\subsubsection{\bf Variability amplitude ($A_{var}$) and the brightness state of source} 
\smallskip
In order to find out whether extent of variability has any dependence on the brightness of the source, we calculated the amplitude of variability ($A_{var}$)  in R-band for all the nights monitored for long enough time to show a minimum of 3\% amplitude variation (see Equation~ \ref{eq4}).  The values of  $A_{var}$  are plotted against nightly averaged brightness magnitudes in R-band (Figure 3) for the duration of 2013 January to 2015 June. We notice larger variability amplitudes when the source was brighter. In blazars, the $A_{var}$  is indicative of the environment where turbulent plasma in the jet interacts with frequent shock formations where relativistic electrons are accelerated in the magnetic field which then cools down  leading to synchrotron radiation.  During this period (2013-2015), the source was in a relatively more active phase showing average R-band magnitude of 13.22 $\pm$ 0.01mag (historical average R = 14.0) and therefore one would expect larger amplitude of variation in the active jet. When the source is relatively faint, thermal emission from the host galaxy is expected to dilute the intrinsic variation in the  jet emission,  resulting in smaller $A_{var}$. Several authors have reported a  similar behaviour to what we have noticed. \citet{agarwal2016} and \citet{yuan2017} notice a very mild trend of larger $A_{var}$ when the source was brighter. \citet{montagni2006} estimated rates of magnitude  variation for 102 nights during 1996--2003 for S5 0716+71 and found faster ($\sim 0.08/h$) rates when source was brighter ($R < 13.4$), though the dependence was weak, compared to average rate of change ($\sim 0.027 mag/h$) irrespective of the state of the source brightness.  

 However, just the opposite  behaviour has been detected  by \citet{kaur-3c2017} in another IBL, 3C66A,  i.e., larger amplitude of variability when the source was relatively fainter. Similar trend was reported by \citet{Chandra2013th, Baliyan2016} in a long-term (2005- 2012) study on the blazar S5 0716+714, reporting larger amplitude of variation when source was relatively fainter. Perhaps more extensive study on several blazars is needed to address this issue.

The behaviour of amplitude of variation as a function of the source brightness  also provides a clue to how the LTV and INV could be related. When the source is bright, it indicates that the  relativistic shock is propagating through the larger scale jet leading to enhanced flux at the longer time scale (LTV) \citet{Romero1999}. The interaction of the shock with local inhomogeneities (small scale particle or magnetic field irregularities) or turbulence interacting with the shock \citep{marscher1992} is perhaps giving rise to the intra-night variations (INV). Since we notice an increase in the amplitude of  INV with an increase in the  mean brightness of the source, later being seen due to LTV, there is perhaps a relation between LTV and INV.  A statistical study on a number of sources with good quality long-term data showing INV, STV and LTV on a large number of nights would, perhaps reveal whether INV amplitudes really have any correlation with the long and short term variability amplitudes. Certainly, S5 0716+714 would qualify as one such candidate for the study. 

\begin{figure}
\includegraphics[width = 0.4\textwidth]{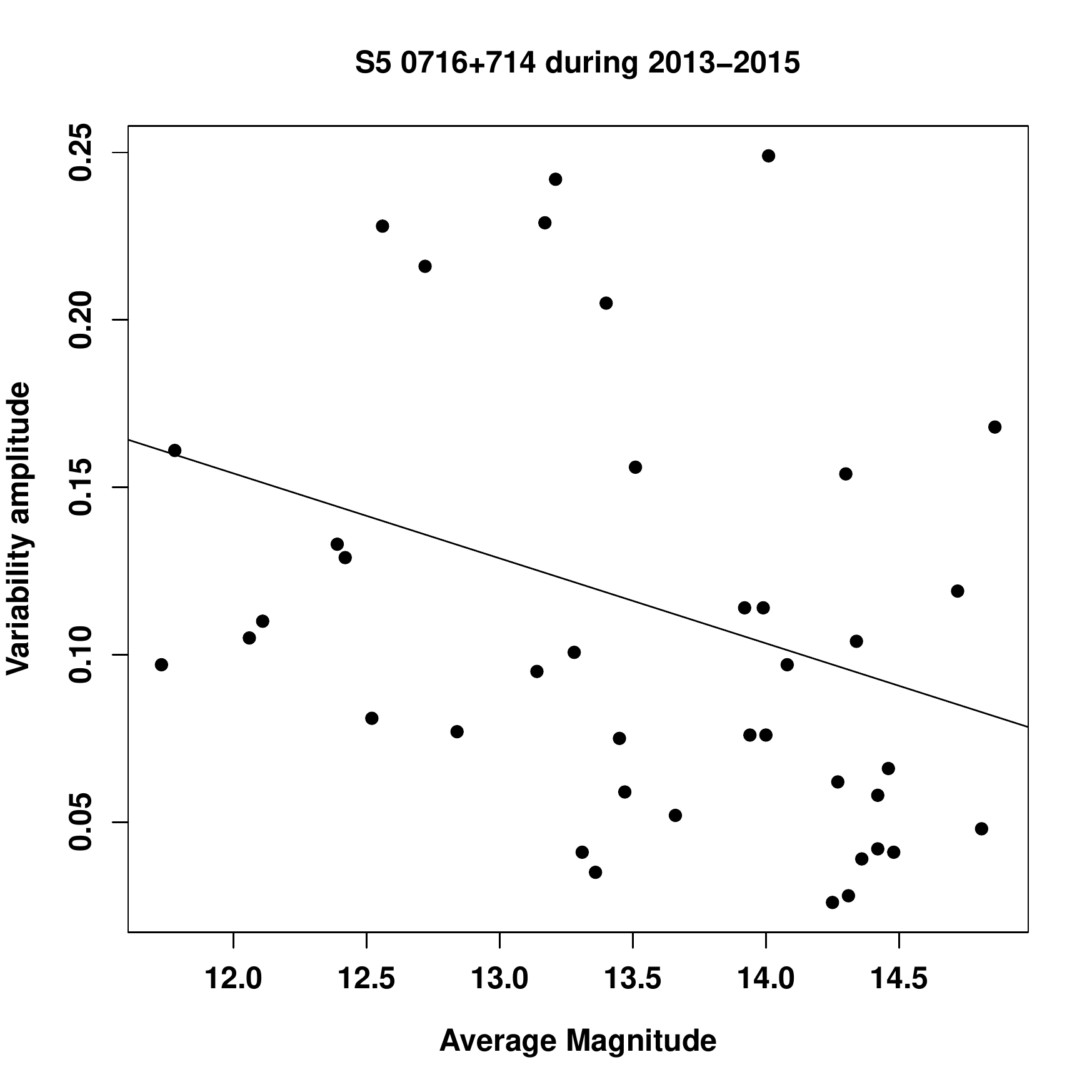}
\caption{The amplitude of variability as a function of the average R-band brightness during 2013-2015.}
\label{ampvar}
\end{figure}

\subsection{Long-term variability}
\begin{figure*}

\includegraphics[width = 0.8\textwidth]{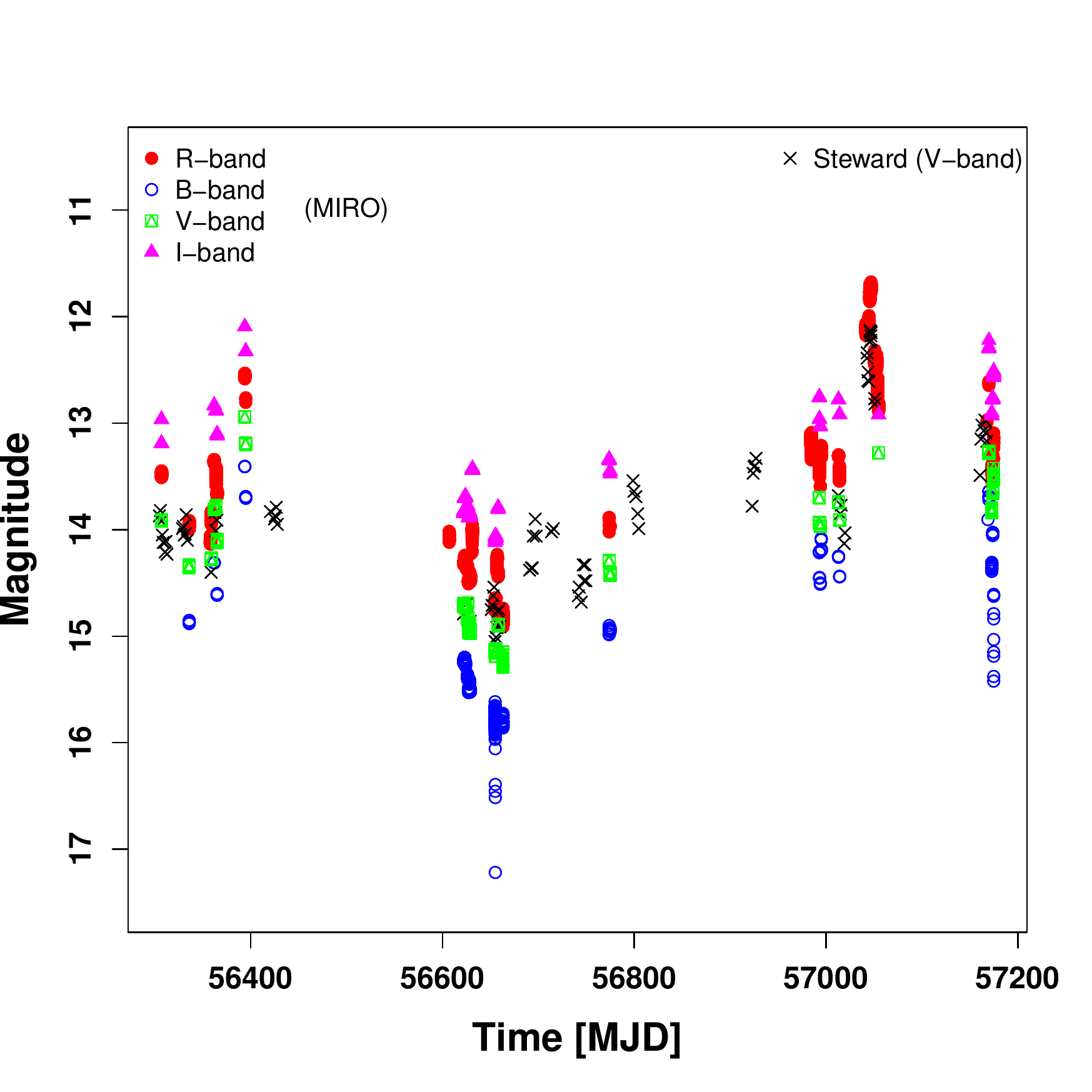}
\caption{Long-term B, V, R \& I band light curves of S5 0716+714 for the duration 2013 January - 2015 June. Data used are from MIRO and Steward observatory. The source has undergone in the brightest and the faintest phases, during 2013$--$ 2015, exhibiting R-band magnitude of 11.68(0.05) and 14.85(0.06) respectively.}
\label{long}
\end{figure*}

The long-term optical light curve constructed for the period 2013 January-2015 June for the S5 0716+714 is shown in Figure 4, with time in MJD and B,V,R \& I brightness in magnitude.  A total of  46 nights  with  6256, 159, 214, \& 177  data points in R, B, V, \& I-bands, respectively, are used in generating these $LC$s. The source was in its faintest state with 14.85(0.06) mag in R-band on MJD 56663.02 (2014 Jan 6)  and in its brightest state with 11.68(0.05) mag, almost one year later on MJD 57040.90 (2015 January 18). 
 S5 0716+71 has undergone several outbursts and flares during its two \& half year journey with two major outbursts peaking in 2013 March and 2015 January, having a duration of about 350 \& 510 days, respectively (see Figure \ref{long}).
 An outburst here is defined as a significant (more than one magnitude) enhancement in the flux over a considerable duration- tens of days to a few months or longer. In our case, limited by the observational data (ours and those from Steward Observatory), we have estimated the outburst duration, looking at the trend of long term variation, based on the above criterion. These outbursts are superposed by a number of fast flares.
We estimate long-term variability (LTV) amplitudes of  about 2.5 mag and 3.45 mag  with time scales of 250 and 360 days, respectively, during these two outbursts. These LTV time scales are estimated with respect to the minimum and maximum brightness values of the source during the two outbursts. During the 2015 January  outburst, S5 0716+714  reached its unprecedented brightness level \citep{atel2015, chandra2015}.  Using multi wavelength data from Fermi-LAT, Swift-XRT, Swift UVOT, MIRO (optical R-band), Steward optical R-band and polarization data, we \citep{chandra2015} detected two sub-flares contributing to this major 2015 January outburst. In optical, the source brightened by 0.8 mag in 6 days (MJD 57035--57041) and, post flare,  decayed over next four days at a rate of ~0.13 mag per day. Very sharp drop in brightness within a day (MJD 57040--57041) and subsequent rise in brightness the very next day (MJD 57042) indicated to the  presence of two sub-flares with almost same peak flux during the outburst. A rapid swing in the position angle of polarization indicated to the magnetic reconnection \citep{zhang2012} in the emission region, causing the outburst. 
 
In the long term, the source became fainter within a year from its average brightness, R = 13.5 mag in 2013 to R $\sim$ 15 mag in 2014 January. It was then in the brightening phase  during  2014 to 2015 with intermittent  flaring activity. S5 0716+71 attained brightest flux value in 2015 January and started its journey towards fainter side later as reflected in all the B, V, R, and I-band (from R = 11.68 mag to 13.20 mag, a 1.52 mag decay in five months, c.f., Figure \ref{long}) $LC$s.  In addition to major outbursts, there are at least 9-flares with their duration ranging from 20 days to 30 days leading to changes in brightness of the source  from a few tenths of magnitude to as much as more than 1.5 magnitude in R-band. The frequent large gaps in the data restrict us from appropriately characterizing these flares which indicate that the source remains almost always active with substantial brightness changes.  During our observation period, S5 0716+714 was brightest on 2015 January 18 (MJD= 57040.90) with a value R= 11.68$\pm0.05$ and faintest on 2014 January 06 (MJD=56663.02) with a value R=14.85$\pm 0.06$.

There are several approaches to explain the variation at various time scales- long as well as short. The intrinsic variations could be caused by the instability \& hot spots in the disk or its outflow \citep{kawaguchi1998, Chakra-wiita1993}, and activities in the relativistic jet\citep{MarscherGear1985, marscher2008}. Variations could also be caused by the processes extrinsic to the source, e.g., interstellar scintillation- which are highly frequency dependent and normally affect long wavelength radio observations, gravitational microlensing - might cause long-term variations in some source but will result in achromatic, symmetric lightcurves. The later is less likely to cause INV \citep{wagner1995}. Since S5 0716+714 was in a relatively bright phase and emission is strongly jet dominated, most probable source of variation should be processes in the jet. The shock-in-jet model \citep{MarscherGear1985, marscher2008} is normally able to explain a variety of variability events with some modifications \citep{zhang2015, Camenzind1992}, where a shock propagates down the jet interacting with a number of particle over-densities or stationary shocks/cores distributed randomly in the parsec scale jet. Such standing shocks are formed due to pressure imbalance between jet plasma and inter-stellar medium (ISM). In trying to maintain a balance, an oblique shock is created perpendicular to the jet axis. The relativistic shocks interacting or passing through such regions energize the particles in the presence of magnetic field, which then radiate synchrotron emission while cooling. Either the jet moves in a helical motion or the blob moves in a helical magnetic field,  causing a change in the viewing angle, thus changing the Doppler factor which significantly enhances/reduces the intrinsic flux variation, depending upon the decrease/increase in the angle. The model can explain rapid variations by resorting to  a jet-in-jet scenario\citep{zhang2015, Camenzind1992}.  The small flares before the outburst indicate  acceleration/cooling of the relativistic particles due to plasma blobs interacting with shock front and an ongoing activity in the jet due to the superposition of all the events leading to long-term variability in the source. In our study, after the second outburst, the source enters into a faint state again.

Several studies have been carried out to address the long-term behaviour of the source. \citep{Raiteri2003, nesci2005} used the historical data, including from the literature, during 1953--2003 and noticed alternate trends of decreasing and increasing mean brightness on a tentative period of about 10 years, claiming precession of the jet to be  responsible for them. It should be noted that even during these slow trends of increasing/decreasing mean brightness levels, source was very active with a large number of flares superposed on the longer trends. Again, a decreasing trend was noticed by \citet{Chandra2013th} beginning 2003 which continued upto 2012, they also predicted an increase in average brightness after 2014.  \citet{agarwal2016} observed the source for 23 nights during 2014 November-- 2015 March and found the source in bright state, showing INV for 7 out of 8 nights and STV with 1.9 mag change during about 28 days (MJD 57013.86 -- 57041.34). In their long-term work for the period 2004 -- 2012, \citet{dai2015} reported STV at 10 days time scale, 11 INV nights out of 72 nights observed with an average magnitude of R=13.25 and an overall change by 2.14 magnitude.

A complete observation log along with daily averaged R band photometric magnitudes for S5 0716+714  and nature of the night, are provided in Table \ref{t-comp}.

\begin{table*}
%\label{t-comp}
\caption{Observation log and photometric results for  S5 0716$+$714 in R-band during 2013 January-2015 June. Columns are: Date of observation, Time (UT) \& MJD, No. of data points, average magnitude with standard deviation  $\&$ Photometric errors, variable (Y/N)}
\textwidth=7.0in
\textheight=10.0in
\vspace*{0.5in}
\noindent
\begin{tabular}{cccrcrrc} 
\hline 
\hline    \nonumber
Date	 & $T_{start}$ & MJD   & N\footnote{Number of data frames} & $\bar{m}$  & $m(\sigma)$ & $E_{phot}$   &Variable  \\
(dd-mm-yyyy) & (hh:mm:ss) &   &  &(Avg mag) &  mag& mag &(Y$/$N)\\
\hline
14-01-2013  	&21:30:14   &56306.89600	 &785  &13.478  	&0.01	&0.01  & N \\
12-02-2013 	&20:10:09	&56335.84038	&195	  &13.943  	&0.02	&0.01	 & Y\\
06-03-2013  	&23:06:00	&56357.96250	&237	  &14.007	&0.10	&0.03  & Y\\
07-03-2013  &00:03:50	&56358.00266		&203  &13.996  &0.10    &0.01  &Y\\
10-03-2013  &19:37:50   &56361.81794 	       &231  &13.356   &0.01	&0.003  &N \\
12-03-2013  &20:14:02	&56363.84308    	&229	  &13.505	&0.05	 &0.004  &Y\\
13-03-2013  	&23:38:51 	&56364.98531	&160  &13.660	&0.01	 &0.005  &N\\
11-04-2013  &19:49:09	&56393.82580		   &182  &12.563	&0.03	 &0.02  &N\\	
12-04-2013  &19:46:43	&56394.82411		   &5  &12.780	&0.02	 &0.02  &N\\	
11-11-2013 	&00:44:51	&56607.03115	&139	  &14.078	&0.02	 &0.003  &Y\\
26-11-2013 	&03:20:25	&56622.13918	&13 	  &14.310	&0.01	 &0.003  &N\\
27-11-2013  &01:48:39	&56623.07545		&25	  &14.259	&0.01	 &0.01  &N\\
28-11-2013  &01:13:25 	&56624.05098		&169  &14.275	&0.01 	&0.01   &N\\
29-11-2013 	&01:07:09	&56625.04663	& 247 &14.337	&0.01	 &0.01  &N\\
30-11-2013  &02:29:12 	&56626.10361 		&107	  &14.357	&0.01	 &0.008  &N\\
01-12-2013  &03:08:06	&56627.13063		&112  &14.485	&0.01	 &0.02  &N\\
02-12-2013  &02:32:13	&56628.10571		&200  &14.417	&0.01	 &0.04  &N\\ 
03-12-2013  &02:12:03   &56629.09170		&242	  &14.463	&0.01	 &0.03  &N\\
05-12-2013  &00:19:24	&56631.01347		&230  &14.011	&0.03	 &0.02  &N\\
28-12-2013  &21:20:00	&56654.88889		&284	  &14.718	&0.02	 &0.007  &Y\\
30-12-2013  &21:43:43	&56656.90536		&350	  &14.304	&0.04	 &0.005  &Y\\
01-01-2014  &00:18:37	&56658.01293		&183	  &14.423	&0.01	 &0.004  &N\\
05-01-2014 	&02:10:57	&56662.09094	&50	  &14.816	&0.01	 &0.005 &N\\
06-01-2014  &00:43:03	&56663.02990		&349	  &14.855	&0.06  &0.006   &N\\
26-04-2014  &20:26:39   &56773.85184		&10	  &13.924	&0.05	 &0.006 &N\\    
27-04-2014  &20:08:22	&56774.83914		&05	  &13.969	&0.01	 &0.008 &N\\
22-11-2014	&20:08:34   &56983.83929	&49	  &13.143	&0.02	 &0.002  &N\\
23-11-2014  &18:47:49	&56984.78321		&207  &13.212	&0.06	 &0.005 &N\\
02-12-2014  &01:13:55	&56993.05133	&284  &13.404	&0.06	 &0.01 &Y\\
03-12-2014  &01:16:34	&56994.05317		&454  &13.276	&0.04	&0.006  &Y\\
22-12-2014  &01:41:07	&57013.07022	&579	  &13.456	&0.08	 &0.01 &N\\
18-01-2015  &18:15:04     &57040.90131  &105	 &11.681  &0.05  &0.05 &N\\
19-01-2015  &16:32:33  	&57041.68927	&442	  &12.114	&0.02	 &0.006	 &N\\
20-01-2015  &18:49:11	&57042.78417		&06	  &12.087	&0.04	 &0.002  &N\\
22-01-2015  &14:42:59	&57044.61319	&100  &12.063  &0.02	 &0.004 &N\\
23-01-2015  &16:37:47	&57045.69292	&934	  &11.776  &0.03	 &0.004  &N\\
24-01-2015  &22:06:24	&57046.92112	&240	  &11.727	&0.02	 &0.01 &N\\
28-01-2015  &17:19:42	&57050.72202		&557	  &12.398	&0.02	 &0.002 &N\\
29-01-2015  &19:03:20	&57051.79399	&401	  &12.518	&0.01	 &0.01 &N\\
30-01-2015  &15:40:12	&57052.65292	&513	  &12.416	&0.02	 &0.01 &N\\
31-01-2015  &16:35:46	&57053.69152	&727	  &12.726	&0.05	 &0.01 &N\\
01-02-2015  &20:38:28	&57054.86005 	&44	  &12.837   &0.02	 &0.01 &N\\
25-05-2015  &21:49:50	&57167.90961	&02	  &12.979	 &0.07	&0.005 &N\\
27-05-2015  &20:35:16	&57169.85782		&05	  &12.626	 &0.01	&0.02 &N\\
30-05-2015  &20:38:38	&57172.86016	&10	  &13.449	 &0.04	&0.01 &N\\
31-05-2015  &20:22:47	&57173.84916		&15	  &13.315	 &0.01	&0.003 &N\\
01-06-2015  &20:13:27	&57174.84267		&18	  &13.171	 &0.05	&0.003 &N\\
\hline
\end{tabular} 
\label{t-comp}
\noindent
\end{table*}

%\begin{table*}
%\label{t3}
%\centering
%\caption{Brightest and faintest magnitudes of the S5 0716+714 during two \& half year period.}
%\begin{tabular}{@{}cccccc@{}}
%\hline
%\hline
%Source	&\hspace{0.35mm}Brightest &\hspace{0.35mm}MJD 		&\hspace{0.35mm} %Faintest		&\hspace{0.35mm}MJD   \\
  %      	&\hspace{0.35mm}	mag	   &\hspace{0.35mm}(Date) 	& \hspace{0.35mm} mag 	%		&\hspace{0.35mm}(Date)\\
%\hline
%S5 0716+714		&\hspace{0.35mm}11.68			&\hspace{0.35mm}57040.90		&\hspace{0.35mm}14.85			&\hspace{0.35mm}56663.02\\
%         		&\hspace{0.35mm}(0.05)			&\hspace{0.35mm}(18-Jan-2015)	&\hspace{0.35mm}(0.06)			&\hspace{0.35mm}(06-Jan-2014)\\        			
%\hline		
%\end{tabular}
%\end{table*}

\subsubsection{Spectral behavior of S5 0716+71}

The variation of color with the brightness of the source provides useful clues to constrain the blazar emission models \citep{hao2010}. To investigate the spectral behavior of  S5 0716+714 over long timescale i.e., from 2013 to 2015, the color-magnitude diagrams,  (B-R) v/s R,  (B-V) v/s V and (V-R) v/s R, are plotted using  nightly averaged magnitudes in B, V and R bands, respectively. The minimum and maximum values of the colour indices for better sampled case of B-R v/s (B+R)/2 are, 0.40 and 1.3, respectively, while the color average is $<B-R>$ = 0.6 mag, with standard deviation $\sigma$=0.14 mag.   
\smallskip
 The Figure 5 shows the spectral behavior of the source with the brightness  during 2013$--$2015. The first panel (from top) shows the (B-R) spectral color versus its average magnitude. Similarly, the middle and the bottom panels display the (B-V) and (V-R) spectral behavior with their average brightness magnitudes, respectively. 
To quantitatively determine the correlation between the color index  with brightness in Figure 5 (Color v/s magnitude),  we performed regression analysis by fitting a straight line, y $=$ mx $+$ c (y = color index, x = average magnitude) using linear model in R software package and extracted various parameters, such as,  intercept(c), slope(m), correlation coefficient(r), p-value etc. The values for these parameters are given in Table \ref{tabcol}. 

\begin{table*}
%\label{tabcol}
\centering
\caption{The values of the regression parameters for color indices  as a function of brightness for S5 0716+714 during 2013-15.}
\begin{tabular}{cccccc}
\hline
\hline
Color Index	& m 	& c	&$r^2$	&r &p  \\
\hline
B-R &0.08 $\pm$ 0.03		&-0.22 $\pm$ 0.45	&0.26	&0.51	& 0.02     \\
B-V	&0.02 $\pm$ 0.02		&0.16 $\pm$ 0.34		&0.06	&0.25	& 0.29     \\
V-R	&0.05 $\pm$0.03		&-0.37 $\pm$ 0.48	&0.12	&0.35	& 0.13     \\
\hline    
\end{tabular}
\label{tabcol}
\noindent 

m = Slope of regression line, $r^2$ = square of Pearson correlation coefficient, p = Probability for null hypothesis.
\end{table*}

\begin{figure}

\includegraphics[width = 0.35\textwidth]{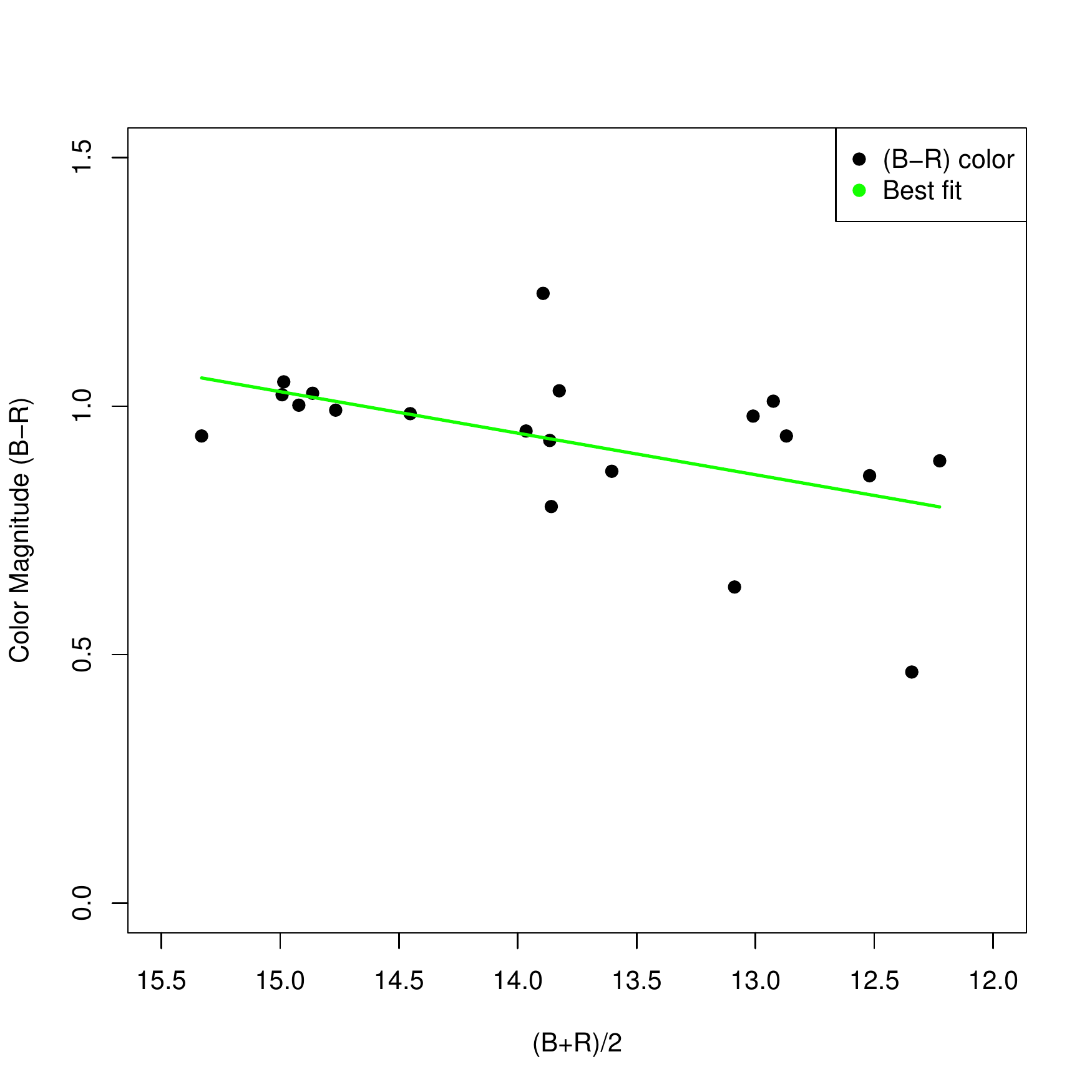}
\includegraphics[width = 0.35\textwidth]{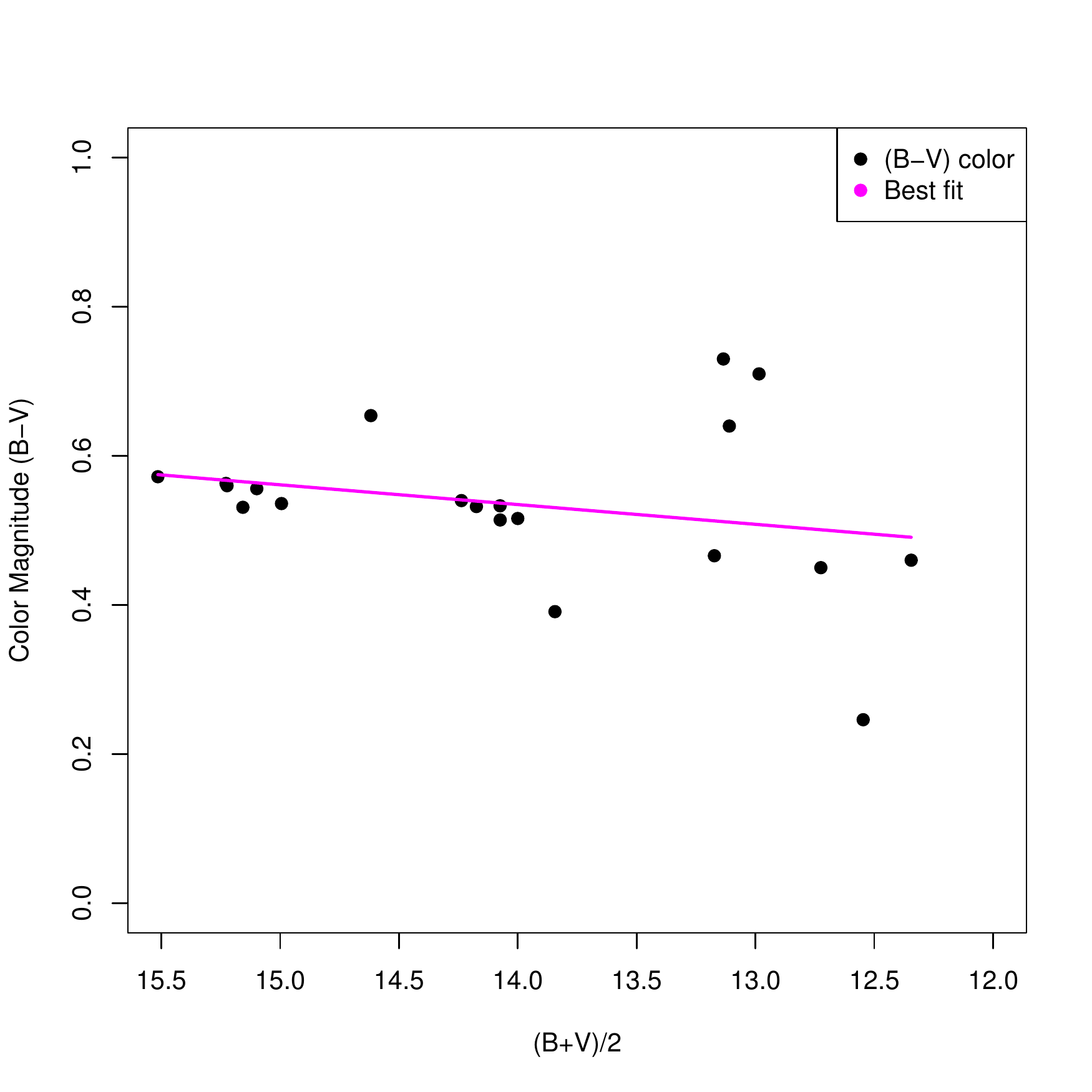}
\includegraphics[width = 0.35\textwidth]{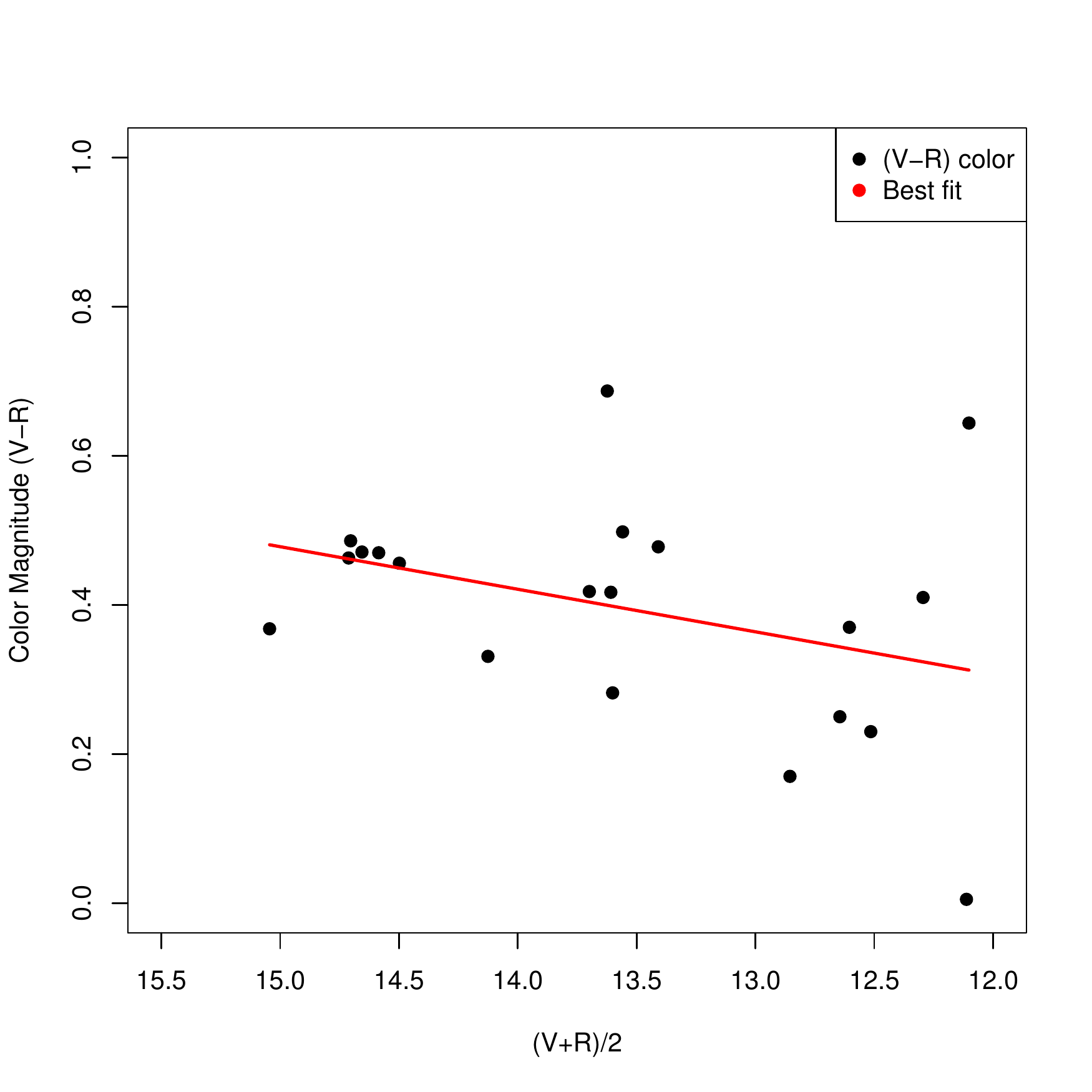}
\caption{Color-magnitude plot for the source S5 0716+714 during 2013-2015, showing bluer when brighter trend. The fit is obtained by performing the linear regression analysis and values for the parameters are given Table \ref{tabcol}.}
\label{allcol}
\end{figure}

It is clear from the Figure \ref{allcol} and values of various parameters obtained from regression analysis (see Table \ref{tabcol}) that the source showed weak positive correlation for B-V and V-R color indices  plotted against brightness magnitudes,  with Pearson correlation coefficient (r) of 0.25 and 0.35 along with p-value of 0.29 and 0.13 respectively. However, comparatively stronger positive correlation for B-R color index versus average magnitude of source with Pearson coefficient $r$ as 0.5 and null hypothesis probability, $p$ = 0.02 are noticed. Thus present study suggests a bluer-when-brighter (BWB) color for S5 0716+71 (cf.,  Figure \ref{allcol})  as also reported by many workers \citep{poon2009, chandra2011, wu2009, man2016}. \citet{li2017} statistically studied the data for S5\, 0716+71 during 1995-2015 and addressed the issue of long-term, short-term and INV behaviour of the symmetry in flares and color and found flares as asymmetric in general  and BWB color on all the time scales considered. The spectral changes in S5 0716+714, and blazars in general,  are complicated and difficult to explain. The source was reported with strong BWB trend over long-timescales \citep{poon2009} and during its flaring phase \citep{ghis1997, wu2005, wu2009, Gu2006}.  \citet{wu2005} and \citet{agarwal2016} discussed color trends in their studies but did not find any change on  the intra-night or long-term timescales.  \citet{Raiteri2003}, on the other hand noticed all possible scenarios, i.e., BWB, RWB and no trend at all,  in their studies. \citet{stalin2009} found source showing no color dependence with brightness on both - the long and short time scales, albeit a BWB color on intra- and inter-night timescales was noticed.  The fresh injection of high energy particles in the emission region inside the jet might lead to BWB behaviour \citep{ghis1997, Raiteri2003, Gu2006}. The BWB behaviour can be explained by  the shock-in-jet model \citep{MarscherGear1985, marscher2008}  where the propagation of a disturbance downstream the jet gives rise to the shock formation and the lag between the emissions at different wavelengths provides information on their relative spatial separations. In the case of  BL Lacs, the higher frequency electrons close to the shock front undergo faster radiation losses than the low-frequency ones. The BWB behavior basically means that the flux enhancements are produced either during the episodes of  intense particle acceleration or, alternatively, by the fluctuating magnetic field superimposed on the local, steady electron energy distribution. The redder when fainter trend indicates that when the jet is not dominant, the contribution from the  disk emission or the host galaxy becomes relevant. These cases show the complex color behaviour of the source with  brightness. However, in case of the S5 0716+714 where host galaxy is several magnitudes fainter,  R \textgreater  20 mag \citep{montagni2006}, thermal contribution from the host is negligible. 

\subparagraph{\bf Spectral variation with time:} 
 \smallskip
The long-term optical light curves of blazars  manifest significant details on the nature of the source as these contain various phases in their brightness, color changes during flares, outbursts  and fainter states. Several authors have looked at spectral variations with time on intra-night and inter-night timescales for blazars \citep{Raiteri2003, wu2005, stalin2009, gaur2012, Rani2010, agarwal2016} and reported a mixed behaviour- some sources showing color dependence while others showed no change in color over considered timescales. \citet{agarwal2016}, during their 130 day study, found a change of about 0.3 mag in spectral color with no significant dependence on the time or brightness phase.  \citet{yuan2017} reported a complex pattern for spectral index with time without any specific trend during the period  2000-2014. The color variations are caused by differential cooling of energetic electrons behind the shock front. The relativistic shock moving down the jet accelerates electrons to high energies at the sites of high magnetic field or electron density giving rise to emission at diverse frequencies. In BL Lacs higher energy electrons cool faster with larger change in  flux with time during a flare.  Since the regions of the plasma over-densities or quasi-stationary shocks are randomly distributed in the jet, the interaction of the relativistically moving knot with existing features in the large scale jet gives rise to multiple outbursts which evolve individually and perhaps differently. The processes involved give rise to  changes in the spectral behaviour with time. 

\begin{figure}
\includegraphics[width = 0.35\textwidth]{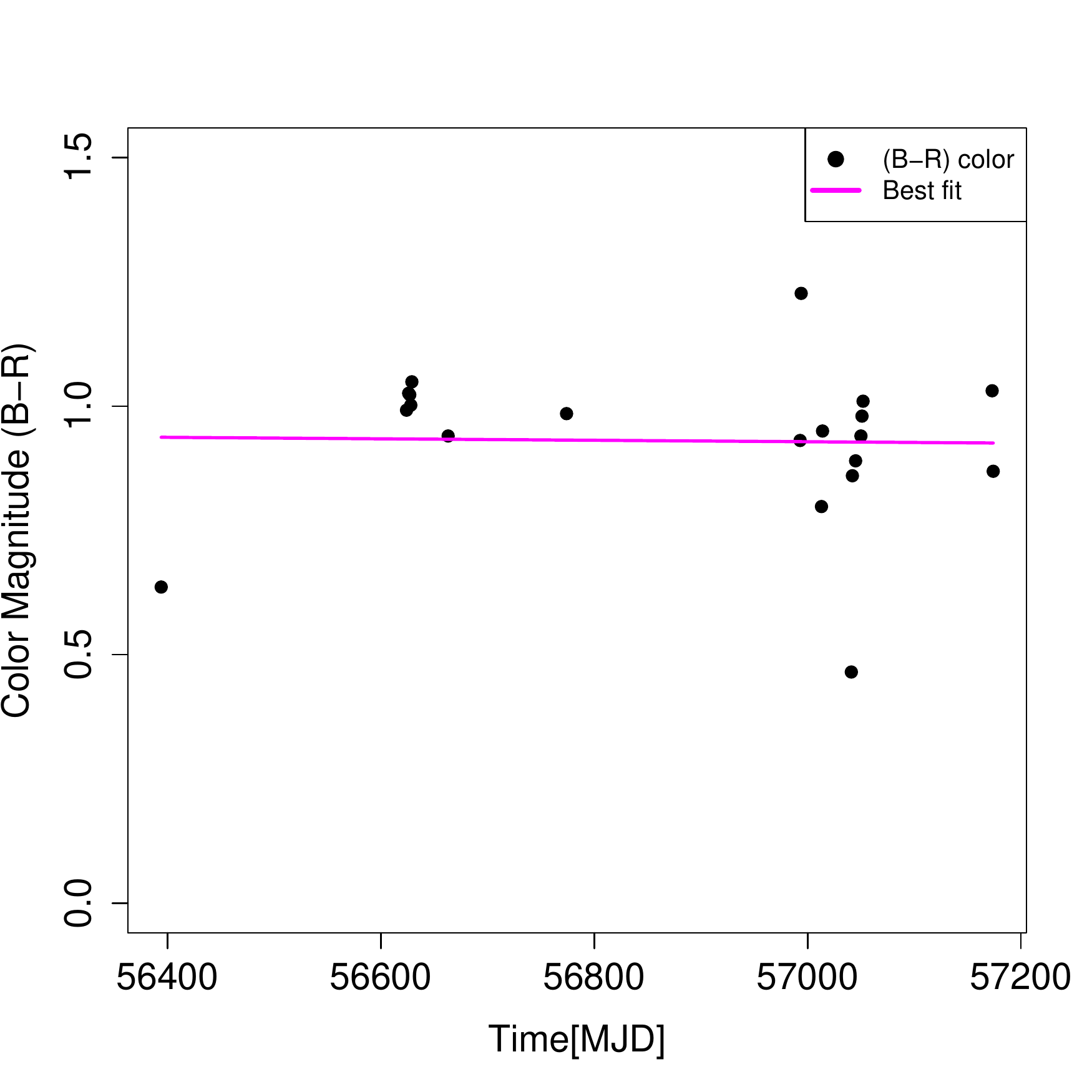}
\includegraphics[width = 0.35\textwidth]{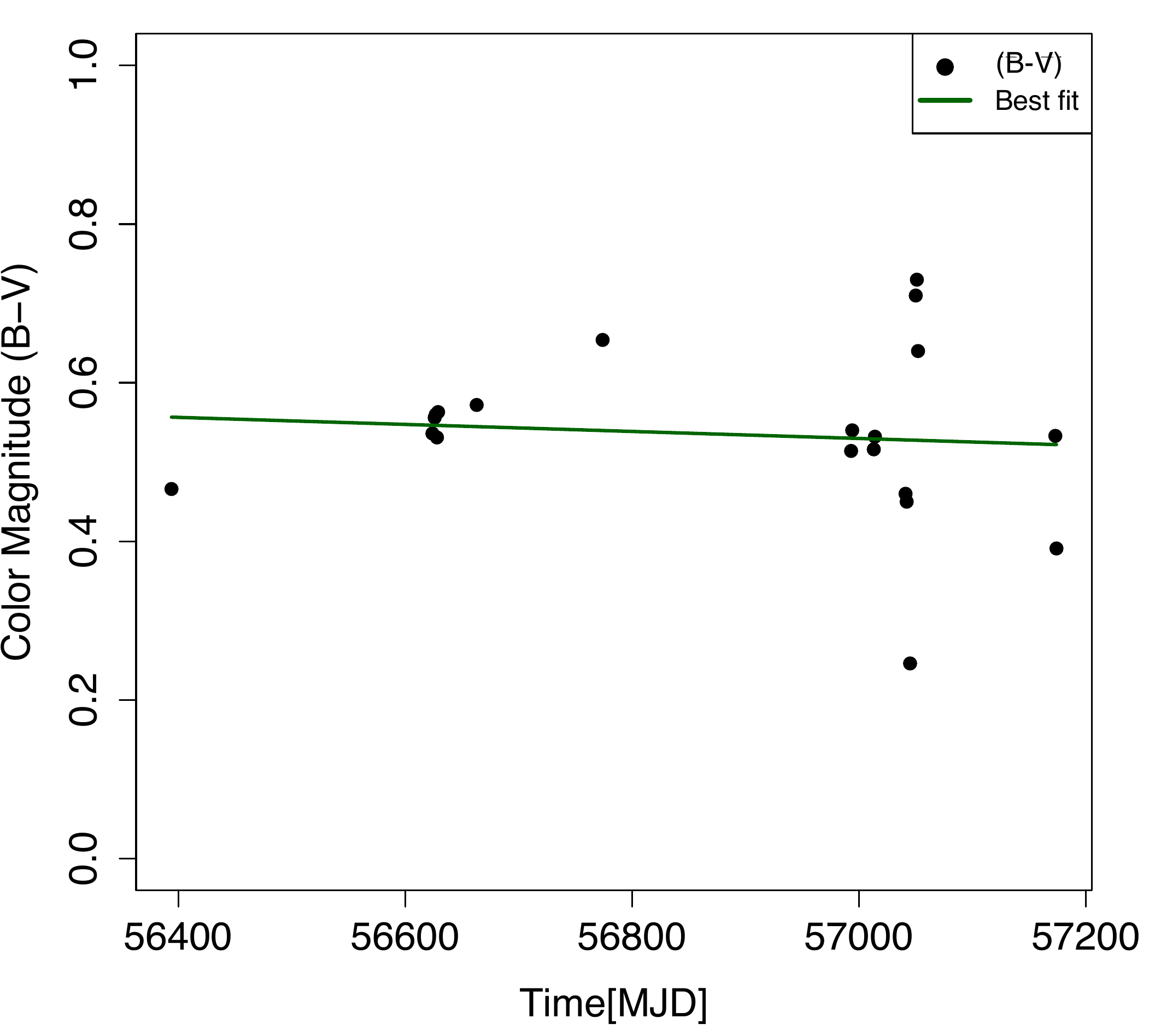}
\includegraphics[width = 0.35\textwidth]{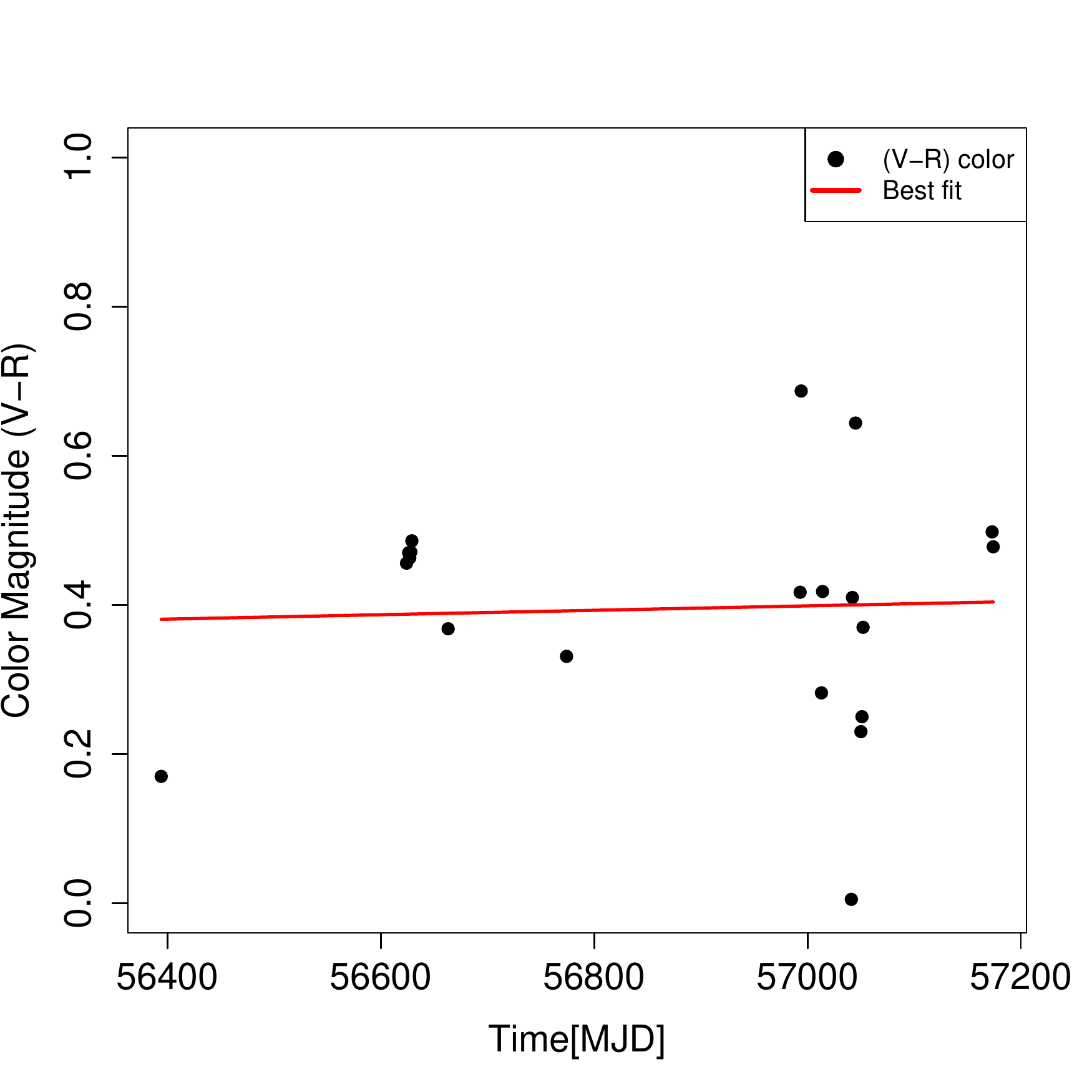}
\caption{The color of the source S5 0716+714 plotted as a function of  time (MJD) during 2013-2015.  The continuous line is the best fit obtained using linear regression analysis of the data.}
\label{coltime}
\end{figure}

 In order to understand the  spectral behaviour of   S5 0716+714 with time (2013 January to 2015 June), we plot color index (B-R), (B-V) and (V-R) against the  time in MJD for this period in Figure \ref{coltime}. To get the correlation between the color index and  time (MJD),  we also performed regression analysis by fitting a straight line, y $=$ mx $+$ c (y = color index, x = Time in MJD ) using linear regression software package and extracted various parameters, such as,  intercept(c), slope(m), correlation coefficient(r) and p-value.  A nicely sampled lightcurve in different optical bands should give a clear picture of the temporal evolution of S5 0716+714. However, our data suffer from substantial gaps and the observations in different bands are not truly simultaneous.   

  The Figure \ref{coltime} shows a mixed behaviour. While we notice a significant bluer spectral  behavior with time in (B-V) color v/s time plot, indicating  the source getting brighter at higher frequencies during this period of two and half years, we see very mild bluer trend in (B-R) v/s Time.  However, the color index  $(V-R)$, shows a mild redder color with time. We, therefore, conclude, based on our data during 2013 January and 2015 June, that the source S5 0716+714 does not show any strong chromatic behaviour, barring (B-V) showing a  bluer behaviour  during this period. This mixed spectral behaviour with time is in line with other studies.  The source was relatively in bright phase during 2013, in low-phase during 2014 and  entered into its brightest phase in January 2015.

\section{Conclusions}
The IBL blazar S5 716 was observed for 46 nights with high temporal resolution during a period of more than two years (2013--2015) in optical BVRI wavelength bands from Mt. Abu InfraRed Observatory (MIRO). It was monitored for more than two hours during 29 nights to address INV. The nightly averaged B, V, R \& I band brightness magnitudes with 6256, 159, 214 \& 177 data points,  were used to discuss long-term variability and color behaviour of the source. The source exhibited intra-night as well as inter-night variability at significant levels. From the present study, following conclusions are drawn: 
\begin{itemize}
\item Source showed variability over diverse timescales i.e., a few tens of minutes to months and a duty cycle of variation of more than 31\%. The DCV appears to be dependent upon monitoring time. Two major outbursts with $\sim$ 370 and 500 days duration superimposed with several flares are noticed. 
\item The structure function analysis leads to the  shortest variability timescale of $~$45.6 minutes, based on which  upper limit on the size of emission region  of the order of $10^{15}$ cm is estimated. There are several time scales longer than this indicating to multi-sized emission regions in the jet. Based on the longest time scale,  the size of emission region is estimated as  $4.8\times10^{15}$ cm.
\item Assuming the  rapid variations to be originated in the vicinity of central engine, black hole mass is estimated to be   5.6$\times10^{8} M_{\odot}$  using shortest variability time scale.
\item The structure function analysis is used to infer a period of about 1.2 hr on the night of 2013 December 28. However, it could easily be red-noise signature as flux enhancements are within 3$\sigma$.
\item The source exhibited a bluer when brighter (BWB) spectral behaviour in the long term $LC$ which supports shock-in-jet model.
\item The brightness of S5 0716+71 shows a mild increase with time during 2013 January--2015 June along with a mild bluer color.
\item A larger amplitude of variation when the source was in relatively brighter state is detected, indicating to synchrotron dominated jet emission. It, perhaps, indicates that long-term and intra-night variabilities are linked.
\end{itemize}
It should be noted that these inferences are drawn from the data with large gaps. However, the data presented here should be very useful for other related statistical and modeling  studies on this very interesting source.

\section{Acknowledgement}
This work is supported by the Department of Space, Govt. of India. We are grateful to the anonymous learned referee for constructive remarks which improved the quality of this work.  We express our thanks to  Mr. Kumar Venkatramani  and past observers as well as MIRO staff  for their help in  observations. We also acknowledge the use the data from the Steward Observatory spectropolarimetric monitoring project which is supported by Fermi Guest Investigator grants NNX08AW56G, NNX09AU10G, NNX12AO93G, and NNX15AU81G \citep{smith2009}. 

{\it Facility:- MIRO:1.2m(PRL-CCD), MIRO:ATVS}

\bibliographystyle{apj}
\bibliography{referencesUnivSub}

\clearpage

\end{document}